\documentclass[a4paper]{appolbprep}
\pdfoutput=1
\usepackage{amsmath,amssymb,amsfonts,amsxtra, mathrsfs, makeidx,graphics,graphicx,amsthm,epsfig,verbatim}
\usepackage[all]{xy}

\newcommand{\bea}{\begin{eqnarray}}
\newcommand{\eea}{\end{eqnarray}}
\newcommand{\ba}{\begin{array}}
\newcommand{\ea}{\end{array}}
\newcommand{\bi}{\begin{itemize}}
\newcommand{\ei}{\end{itemize}}
\newcommand{\ben}{\begin{enumerate}}
\newcommand{\een}{\end{enumerate}}
\newcommand{\bean}{\begin{eqnarray*}}
\newcommand{\eean}{\end{eqnarray*}}
\newcommand{\eref}[1]{(\ref{#1})}

\newcommand{\tref}[1]{Table~\ref{#1}}
\newcommand{\fref}[1]{Figure~\ref{#1}}
\newcommand{\nn}{\nonumber}

\newcommand{\tr}{\mathop{\rm Tr}}

\newcommand{\BC}{\mathbb{C}}

\newcommand{\BP}{\mathbb{P}}
\newcommand{\BZ}{\mathbb{Z}}

\newcommand{\CMm}{{\cal M}^{\mathrm{mes}}}
\newcommand{\gm}{ g^{\mathrm{mes}}}

\newcommand{\CN}{{\cal N}}

\newcommand{\CC}{{\cal C}}

\newcommand{\CP}{\mathbb P}

\newcommand{\firr}[1]{{}^{{\rm Irr}}\!{\cal F}^{\flat}_{#1}}

\def\theequation{\thesection.\arabic{equation}}

\begin{document}
\pagestyle{plain}
\title{{\Large \bf Brane Tilings, M2-branes and Chern-Simons Theories}
%\thanks{Presented at the XLIX Cracow School of Theoretical Physics 2009}
}
\author{John Davey, Amihay Hanany, Noppadol Mekareeya and Giuseppe Torri
\address{Theoretical Physics Group, The Blackett Laboratory \\
Imperial College London, Prince Consort Road\\ 
London,  SW7 2AZ,  UK\\ $~$ \\
 {\tt j.davey07, a.hanany, n.mekareeya07, giuseppe.torri08@imperial.ac.uk}
}}
%\vspace{-1cm}
{\hspace{\stretch{1}}
\texttt{Imperial/TP/09/AH/06}\\
\maketitle
\begin{abstract}
We investigate $(2+1)$-dimensional quiver Chern-Simons theories that arise from the study of M2-branes probing toric Calabi-Yau 4-folds. These theories can be elegantly described using brane tilings. We present several theories that admit a tiling description and give details of these theories including the toric data of their mesonic moduli space and the structure of both their Master space and baryonic moduli space. Where different toric phases are known, we exhibit the equivalence between the vacua. We identify some of the mesonic moduli spaces as cones over smooth toric Fano 3-folds.
\end{abstract}
%\PACS{PACS numbers come here}
  
\section{Introduction}
Recently, there has been substantial progress in understanding M-theory on various different backgrounds. In particular M-theory on backgrounds of $\mathrm{AdS}_4 \times X^7$, where $X^7$ is a Sasaki-Einstein 7-manifold, has been studied in great detail. These geometries are believed to correspond to world-volume theories of M2-branes that probe Calabi-Yau 4-fold singularities \cite{HullEtAl,Plesser,Martelli:2008si,Hanany:2008cd}. These singularities can be identified with the cone over the aforementioned Sasaki-Einstein manifolds.

When M2-branes probe a Calabi-Yau 4-fold that admits a toric description, the branes' world-volume is thought to be well described by a $\CN=2$ $(2+1)$-dimensional quiver Chern-Simons (CS) theory \cite{Martelli:2008si, Hanany:2008cd, Ueda:2008hx, taxonomy, Hewlett:2009bx} which can be elegantly represented by a {\it brane tiling} \cite{Hanany:2008cd, Hanany:2008fj, phase, higgs, Davey:2009bp}. This brane tiling technology was originally developed to understand the $(3+1)$-dimensional gauge theories that describe D3-branes probing toric Calabi-Yau (CY) 3-fold singularities \cite{Hanany:2005ve,Franco:2005rj}, \cite{Davey:2009bp, Hanany:2005ss,Franco:2005sm}, \cite{Kennaway:2007tq, Yamazaki:2008bt}. It is convenient and perhaps not too surprising \cite{Hanany:2008cd, Hanany:2008fj} that the tilings used to describe M2-brane theories have many features in common with the original D3-brane tilings. Brane tilings have proven to be an incredibly powerful tool for studying a number of interesting phenomena, for example transitions between different singularities using the {\it Higgs mechanism} \cite{higgs, unhiggs} and also {\it toric duality} \cite{taxonomy, phase, Amariti:2009rb}. 

In this paper, we summarize an exploration of a class of gauge theories that arise from the study of M2-branes probing CY 4-folds which are cones over smooth toric {\it Fano 3-folds} \cite{Hanany:2009vx}. These Fano 3-folds are 3 dimensional complex manifolds admitting positive curvature.
 It is known that there are precisely $18$ of these surfaces 
\cite{Hanany:2009vx, toricfano3, database}. It is thought that the investigation of Fano 3-folds may be as fruitful as the recent intensive study of their 2 dimensional analogues (the del Pezzo surfaces) \cite{Feng:2000mi, Feng:2001xr, Feng:2002fv} \cite{Franco:2005rj}.

\section{The $\CN=2$ supersymmetric CS theories in $(2+1)$ dimensions}
In this paper we consider brane tilings that correspond to $(2 + 1)$-dimensional $\mathcal{N}=2$ supersymmetric CS theories. Each theory admits a $U(N)^G$ gauge symmetry, has matter fields that transform in bi-fundamental and adjoint representations, and has specific set of interactions. The Lagrangian for such a theory can be written in $\CN=2$ superspace notation as
\begin{align} \label{lagrange}
\mathcal{L} =& -\int d^4 \theta\left( \sum\limits_{X_{ab}} X_{ab}^\dagger e^{-V_a} X_{ab} e^{V_b}
-i \sum\limits_{a=1}^G k_a \int\limits_0^1 dt\; V_a \bar{\mathcal{D}}^{\alpha}(e^{t V_a} \mathcal{D}_{\alpha} e^{-tV_a})
\right) + \nn\\
& \int d^2 \theta \;W(X_{ab}) + \mathrm{c.c.}~,
\end{align}
where $a$ indexes the gauge groups ($a=1,\ldots, G$) and $X_{ab}$ are bi-fundamental chiral superfields, accordingly charged. $V_a$ are the vector multiplets, $\mathcal{D}$ is the superspace derivative, $W$ is the superpotential and $k_a$ are the CS levels, which are integer valued. An overall trace is implicitly taken since all of the fields are matrix-valued. Each chiral superfield appears exactly twice in the superpotential - once in a positive term and once in a negative term. This is known as the {\it toric condition} on the superpotential \cite{Feng:2002zw}.

The vacuum equations are given by
\begin{align}
\partial_{X_{ab}} W &= 0~, \label{e:fflat} \\
\mu_a(X) := \sum\limits_{b=1}^G X_{ab} X_{ab}^\dagger - 
\sum\limits_{c=1}^G  X_{ca}^\dagger X_{ca} + [X_{aa}, X_{aa}^\dagger] &=  4k_a\sigma_a~, \label{DF1} \\
 \sigma_a X_{ab} - X_{ab} \sigma_b &= 0 \ . \label{DF2}
\end{align}
The first set of equations \eref{e:fflat} are the \emph{F-term equations}, whose space of solutions is called the {\it Master space} \cite{master}.
The others - \eref{DF1} and \eref{DF2} are called the \emph{D-term equations} in analogy to the vacuum equations of $\CN=1$ gauge theories in $(3+1)$ dimensions, with the last equation \eref{DF2} being a new addition.  

It should be noted that, in the absence of CS terms, this theory can be viewed as a dimensional reduction of a $(3 + 1)$-dimensional ${\CN}=1$ supersymmetric theory. 
In particular, $\sigma_a$, the real scalar in the vector multiplet, arises from the zero mode of the component of the vector field in the reduced direction.
We refer to the space of all solutions of \eref{e:fflat}, \eref{DF1} and \eref{DF2} as the {\it mesonic moduli space}, and denote it as $\CMm$.  

It can be shown that
\bea
\label{sumk}
\sum_{a} k_a = 0
\eea 
is a necessary condition for the moduli space to have a branch which is a Calabi--Yau four-fold \cite{Martelli:2008si, Hanany:2008cd, Ueda:2008hx}.
This branch can be interpreted as the space transverse to the M2-branes. 

Let us consider the abelian case\footnote{The mesonic moduli space of the non-abelian $U(N)^G$ theory is expected to be the $N$-th symmetric product of the moduli space for the abelian case, even though a direct derivation is still evasive.} in which the gauge group is $U(1)^{G}$.
We consider the branch in which all of the bi-fundamental fields are
generically non-zero.  In this case, the solutions to the first set of equations \eref{e:fflat} give the {\it irreducible component} of the Master space, $\firr{~}$ \cite{master}.

The third equation \eref{DF2} sets all $\sigma_a$ to a single field, let's say $\sigma$.

The second set of equations in \eref{DF1} consists of $G$ equations.  The sum of all of these equations is identically zero, and so there are actually only $G-1$ linearly independent equations. 
These $G-1$ equations can be divided into one along the direction of the vector $k_a$, and $G-2$ perpendicular to the vector $k_a$.
The former fixes the value of $\sigma$ and leaves a $\mathbb{Z}_k$ action, where $k \equiv  \gcd(\{k_a\})$, by which we need to quotient out in order to obtain the mesonic moduli space. The remaining $G-2$ equations can be imposed by the symplectic quotient of $U(1)^{G-2}$.
Thus, the mesonic moduli space can be written as
\bea
\CMm = \firr{} // \left(U(1)^{G-2} \times \BZ_k \right)~. \label{mesonic}
\eea
The reader should note that these $G-2$ directions correspond to {\it baryonic charges} that arise from D-terms although the \emph{total} number of baryonic charges is four less than the number of external points of the toric diagram \cite{phase}.

\section{Brane tilings for M2 branes}
In this work we restrict our attention to how brane tilings relate to M2-brane theories, although the relationship between tilings and the world-volume physics of D3-branes is a fascinating subject.

A brane tiling (or dimer model) is a periodic bipartite graph on the plane. 
Alternatively, we may draw it on the surface of a 2-torus by taking the smallest repeating
structure (known as the fundamental domain) and identifying opposite edges \cite{Hanany:2005ve}.
The bipartite nature of the graph allows us to colour the nodes either white or black such
that white nodes only connect to black nodes and vice versa.

There is a simple dictionary between a tiling and the Chern-Simons theory
that it represents (\tref{dictionary}). If a tiling is to correspond to a Chern-Simons theory, a set of levels, $k_a$ must be specified. A tiling equipped with these levels is enough information to fully reconstruct a quiver Chern-Simons theory's Lagrangian \cite{Hanany:2008cd, Hanany:2008fj}.

\begin{table}[h!]
\begin{center}
\begin{tabular}{|c|c|c|}
\hline
Tiling & Quiver  & Meaning in gauge theory \\
\hline
Face (tile)& Node & $U(N)$ gauge group \\
Edge & Arrow & A bi-fundamental chiral multiplet \\
Node & A closed path$^{*}$ & An interaction term in the superpotential \\
\hline
\end{tabular}
\caption{A brane tiling dictionary.  
$^{*}$It is important to note that although each term of the superpotential corresponds to a closed path in the quiver, \emph{not all} closed paths of the quiver give rise to the terms in the superpotential. White (black) nodes in the tiling correspond to positive (negative) superpotential terms.
}
\label{dictionary}
\end{center}
\end{table}
The tiling and quiver of the well known ABJM model are given in \fref{ABJMqt} as an illustrative example of how the two objects are related to one another.
\begin{figure}[h!]
\begin{center}
  \hskip -6cm
  \includegraphics[totalheight=4cm]{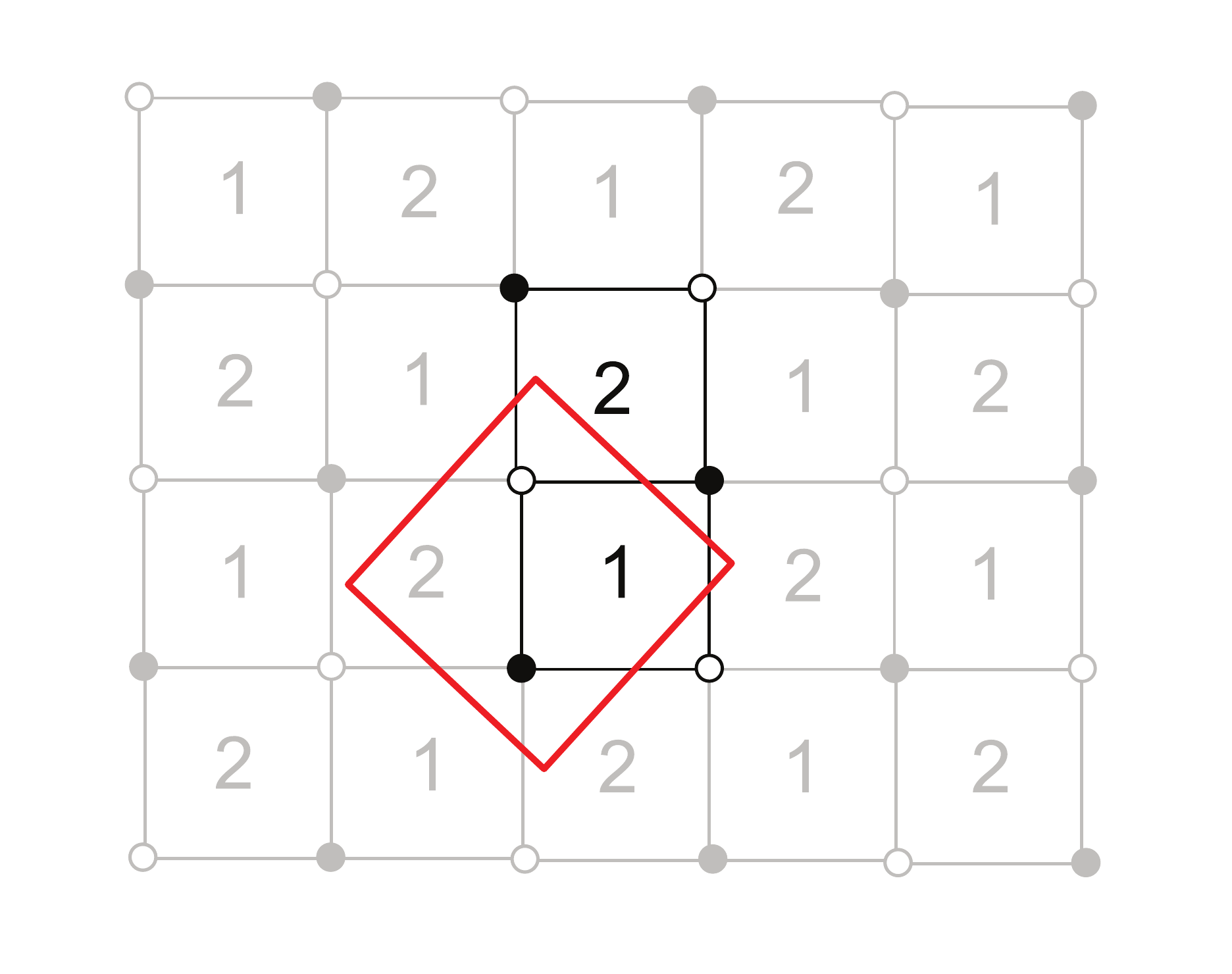}
  \vskip -2.4cm
  \hskip 5cm
  \includegraphics[totalheight=1.2cm]{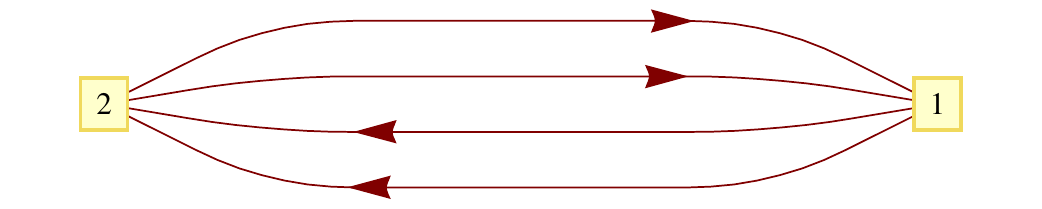}
  \vskip 1cm
\end{center}
\caption{An example of a brane tiling (left) and its corresponding quiver diagram (right). The red square in the tiling indicates the smallest unit of repetition called the \emph{fundamental domain}.  In $(3+1)$ dimensions, this model is known as the conifold theory. 
In $(2+1)$ dimensions, if a CS level $k$ is assigned to one gauge group and $-k$ to the other, then the corresponding model is the ABJM theory.}
\label{ABJMqt}
\end{figure}
\subsection{From a brane tiling to the moduli space}
The brane tiling is a very powerful tool for establishing the relationship between a large class of Chern-Simons theories and their mesonic moduli spaces. In this section we restrict ourselves to the study of abelian Chern-Simons theories corresponding to world-volume theories of one M2-brane.

When a quiver Chern-Simons theory admits a tiling description, we can easily construct the toric diagram of the mesonic moduli space by using the fast forward algorithm which is outlined below:
\begin{enumerate}
\item Assign an integer $n_{{}_X}$ to the edge corresponding to the chiral field $X$ (\fref{f:ph1c4}) such that the CS level $k_a$ of the gauge group $a$ is given by\footnote{This way of representing $k_a$ is introduced in \cite{Hanany:2008cd} and is also used in \cite{Imamura:2008qs}.}
\bea
k_a = \sum_{\text{all fields}~X} d_{a {}_X} n_{{}_X}~, \label{kn}
\eea
where $d_{a {}_X}$ is the charge of the chiral field $X$ under the $U(1)$ gauge group $a$ and can be read off from the quiver diagram easily. 
Due to the bipartite nature of the tiling, we see that the relation $\sum_{a} k_a = 0$ is automatically satisfied.   
\item Define the \emph{Kasteleyn  matrix} $K(x,y,z)$ whose entries are given by
\bea
K_{pq} (x,y,z) = \sum_{\substack{X:~p \leftrightarrow q}} X z^{n_{X}} w_{{}_X}(x,y)~,
\eea
where the summation runs over the edges corresponding to the chiral fields X connecting the node $p$ and the node $q$,
and the weight $w_{_X}(x,y)$ takes the values $x^\alpha y^\beta$ (where $\alpha$ and $\beta$ depend on the orientation of the edge) 
if the edge $X$ crosses the fundamental domain and $w_{_X}(x,y) =1$ if it does not.
\item Take the permanent\footnote{The permanent is similar to the determinant: the signatures of the permutations are not taken into account and all terms come with a $+$ sign. One can also use the determinant but then certain signs must be introduced \cite{Hanany:2005ve,Franco:2005rj}.} of the Kasteleyn matrix. It can be written in the form:
\bea
\mathrm{perm}~K = \sum_{\alpha=1}^c p_\alpha ~ x^{u_\alpha} y^{v_\alpha} z^{w_\alpha}~. \label{permk}
\eea
Each $p_\alpha$, which is a collection of the chiral fields, is called a \emph{perfect matching}.
It is known that the Master space is parametrised by the perfect matchings \cite{master}.
\item The coordinates $(u_\alpha, v_\alpha, w_\alpha)$ of the $\alpha$-th point in the toric diagram are given respectively by the powers of $x, y, z$ in \eref{permk}.  These coordinates can be collected in the columns of the following matrix:
\bea
G_K = 
\left( \begin{array}{ccccc}  u_1& u_2& u_3&\ldots & u_c \\ v_1& v_2&  v_3&\ldots & v_c \\ w_1& w_2& w_3&\ldots & w_c \end{array} \right)~.
\label{e:gk}
\eea
\end{enumerate}
\noindent {\bf Remark 1:} There are redundancies in the $G_K$ matrix.  
In particular, we can construct $\widetilde{G}_K$ (a $(4 \times c)$ matrix) by prepending $(1~1~1~ \ldots ~1)$ into the first row of the $G_K$ matrix.
After performing a series of elementary operations (or equivalently by applying a suitable $GL(4, \BZ)$ transformation) on the rows of $\widetilde{G}_K$ such that the first row is kept to be $(1~1~1~ \ldots ~1)$,
we then remove this first row and obtain another $3 \times c$ matrix $G'_K$.
The matrices $G_K$ and $G'_K$ carry the same toric data, and hence correspond to the same mesonic moduli space\footnote{This arbitrariness in how the fundamental domain was drawn on the tiling contributes to this redundancy.}.
\newline
\noindent {\bf Remark 2:} The $G_K$ matrix contains information about the mesonic global symmetry of the theory. 
In particular, we can transform $G_K$ as stated in Remark 1 so that the rows of the resulting matrix contain weights of the mesonic symmetry. 
\begin{figure}[h!]
\begin{center}
\includegraphics[scale=0.7]{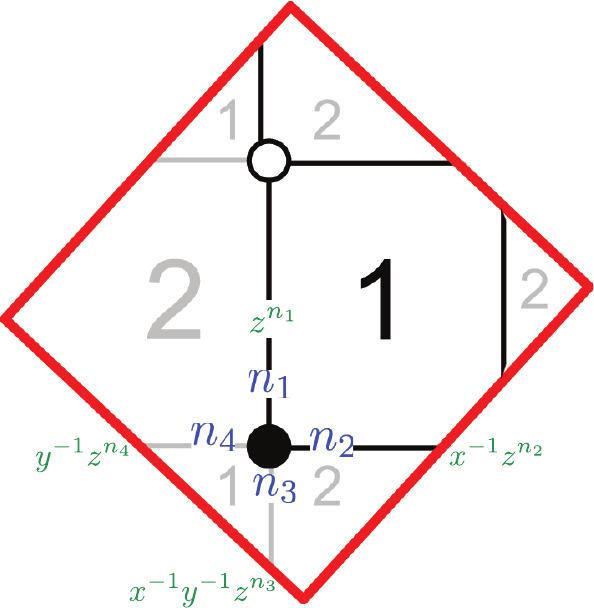}
\caption{The fundamental domain of the tiling of the ABJM theory. Assignments of the integers to the edges are shown in blue and the weights of these edges are shown in green.}
\label{f:ph1c4}
\end{center}
\end{figure}
\section{Toric dualities} 
A toric duality is a situation in which one singular CY variety corresponds to more than one Chern-Simons theory (which we shall refer to as \emph{toric phase}).
Toric phases share several common features, even though their brane tilings are different:
\begin{itemize}
\item The mesonic moduli spaces of all phases are identical.
\item The perfect matchings of different phases are exactly the same (including the labels and up to zero R-charge perfect matchings).  They are charged in the same way under global symmetries.
\item When written in terms of the perfect matchings, the mesonic generators of different phases are precisely the same (up to zero R-charge perfect matchings).
\end{itemize}
Let us now illustrate this idea of toric duality by giving different phases of the $\BC^4$ theory as well as the $\CC \times \BC$ theory.
\subsection{The $\BC^4$ Theory}
There are two known phases of the $\BC^4$ theory:\\

\noindent {\bf Phase I: The ABJM theory with $\vec{k} = (1,-1)$.} 
The quiver and tiling are drawn in \fref{ABJMqt}. In the abelian case ($N=1$), the superpotential of the ABJM theory vanishes, as the chiral fields are simply complex numbers.
Hence, the Master space is $\BC^4$. Since the number of gauge groups is $G=2$, from \eref{mesonic}, it follows that for
the CS levels $\vec{k} = (1,-1)$ the mesonic moduli space is $\BC^4$.  This is parametrised by $X^i_{12}, X^i_{21}$ ($i=1,2$), each of which has an R-charge $1/2$.
$~$ \newline

\begin{figure}[h!]
 \centerline{  \epsfxsize = 4.2cm  \epsfbox{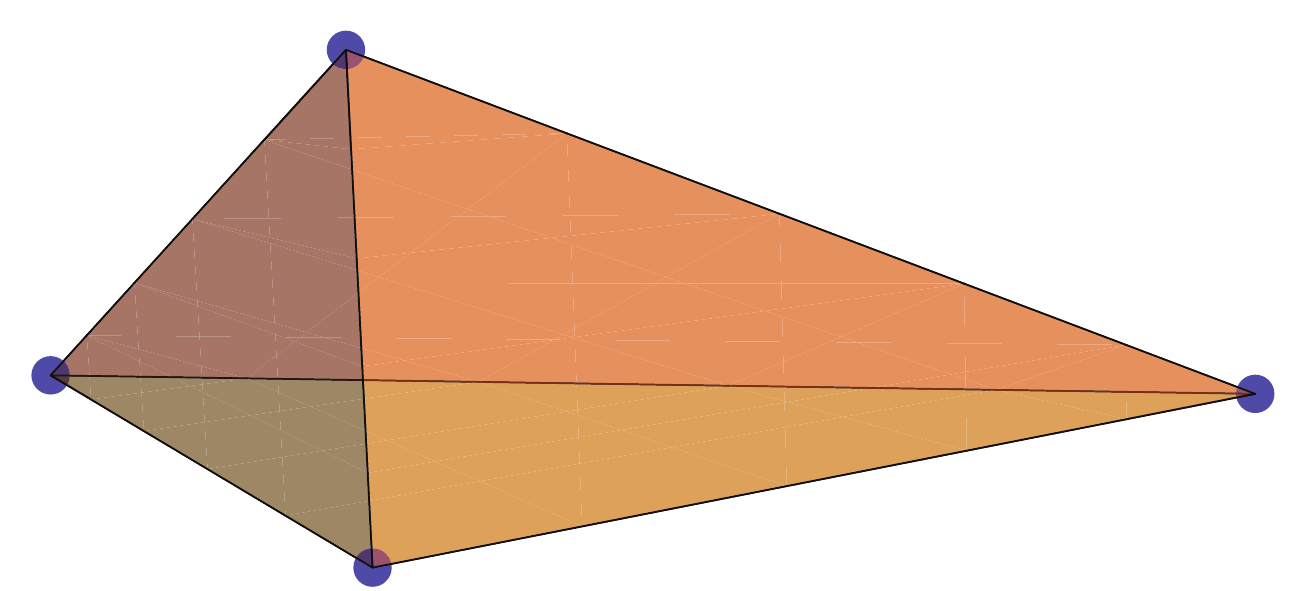} }
 \caption{The toric diagram of the $\BC^4$ theory.}
  \label{f:torabjm}
\end{figure}

\noindent {\bf Phase II: The double bonded hexagon model with $\vec{k} = (1,-1)$.}
The quiver and tiling of this phase of $\BC^4$ is drawn in \fref{ph2c4}.  By a similar argument to the one above, it can be shown that the mesonic moduli space for $\vec{k} = (1,-1)$ is also $\BC^4$ \cite{phase}.
This is parametrised by $X_{12}, X_{21},\phi_i$ ($i=1,2$), each of which has an R-charge of $1/2$.
\begin{figure}[h!]
\begin{center}
  \vskip 0.5cm
  \hskip -7cm
  \includegraphics[totalheight=1.7cm]{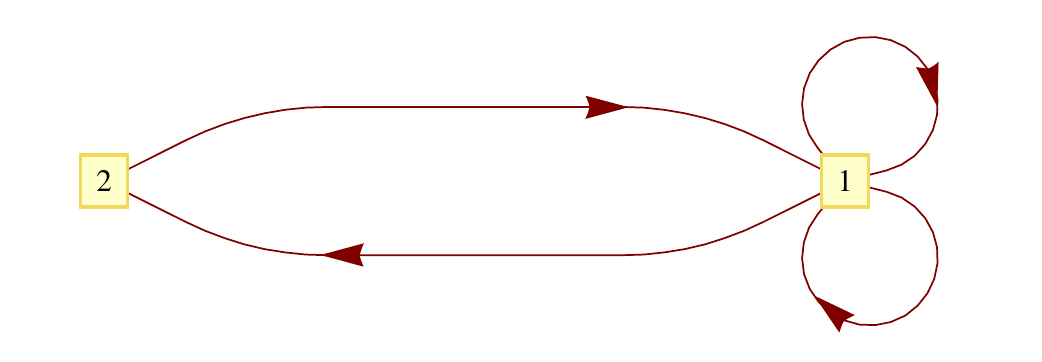}
  \vskip -2.5cm
  \hskip 4cm
  \includegraphics[totalheight=3cm]{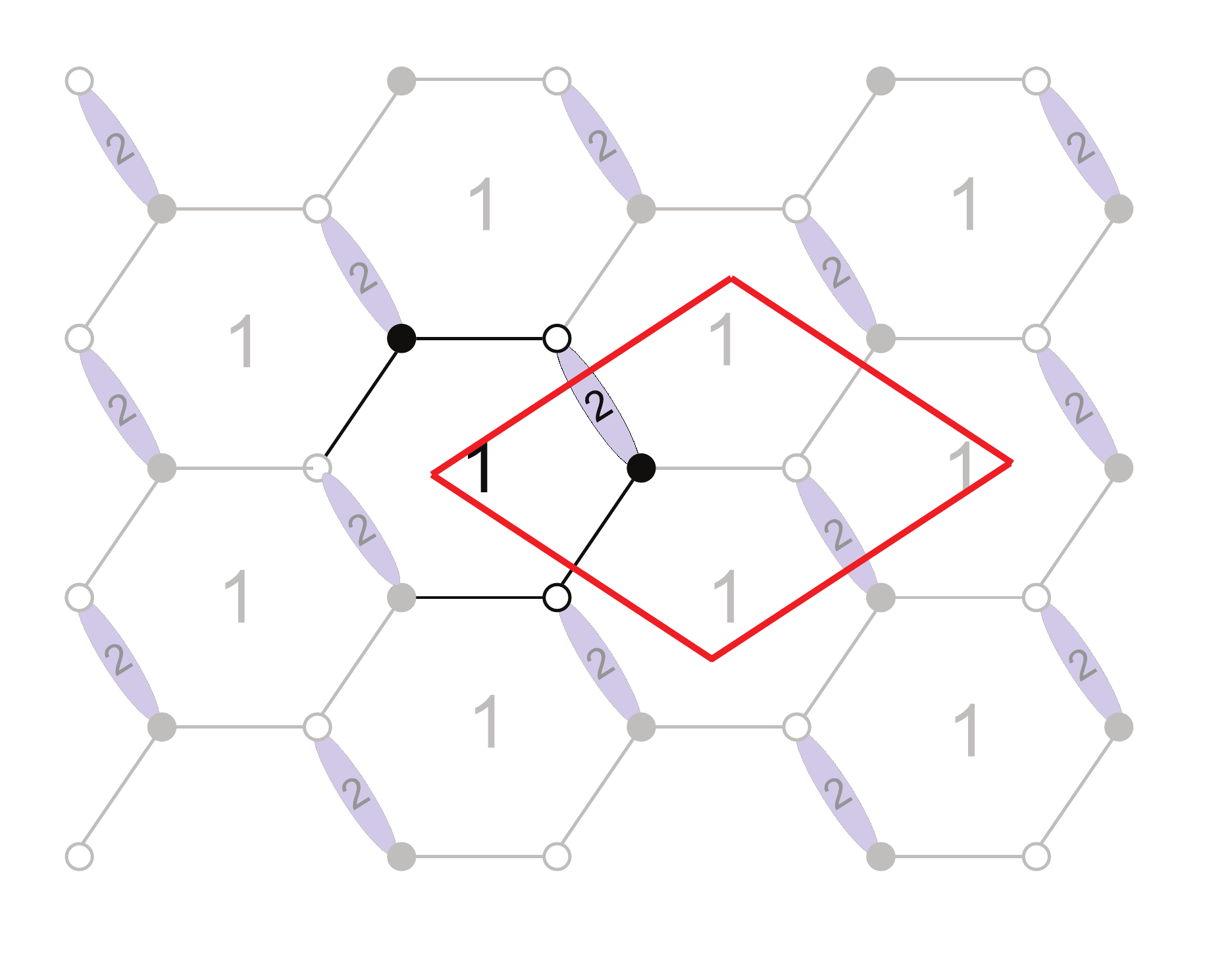}
\caption{Phase II of $\BC^4$.  The superpotential is $W = \mathrm{Tr}(X_{12} X_{21} [\phi_1, \phi_2])$.}
\label{ph2c4}
\end{center}
\end{figure}

\subsection{The $\text{conifold ($\CC$)} \times \BC$ Theory}
There are 3 known phases of the $\CC \times \BC$ theory.  Their quivers and tilings are given in Figures \ref{ph1conxc}, \ref{ph2conxc} and \ref{ph3conxc}.
The toric diagram is in \fref{f:torconxc}.
\newline $~$ \newline
\begin{figure}[htbp]
\begin{center}
  \vskip 0cm
  \hskip -7cm
  \includegraphics[totalheight=3.8cm]{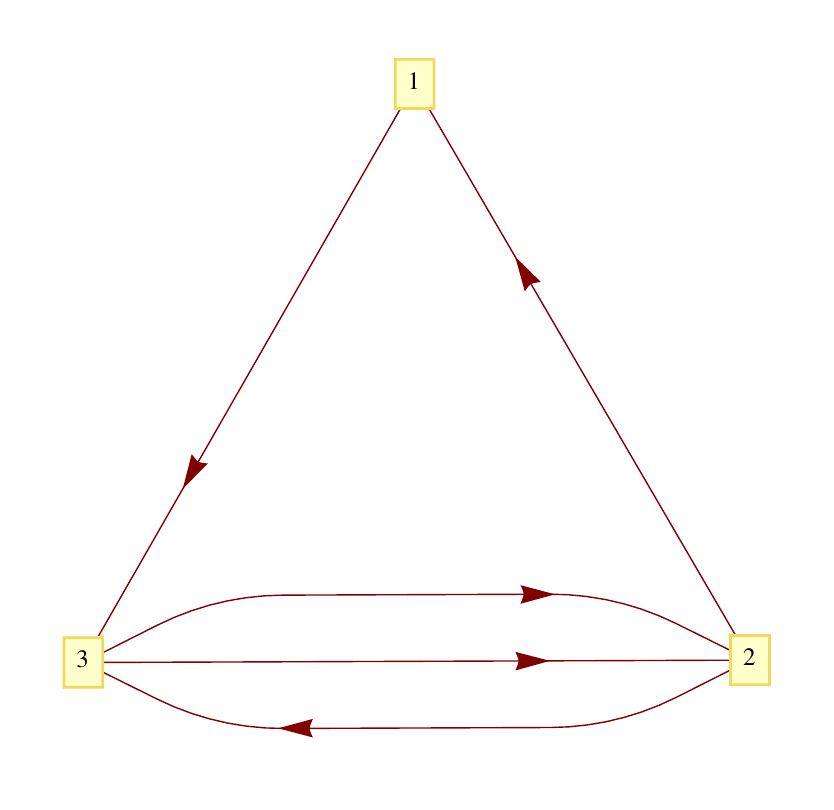}
  \vskip -4cm
  \hskip 4cm
  \includegraphics[totalheight=3.8cm]{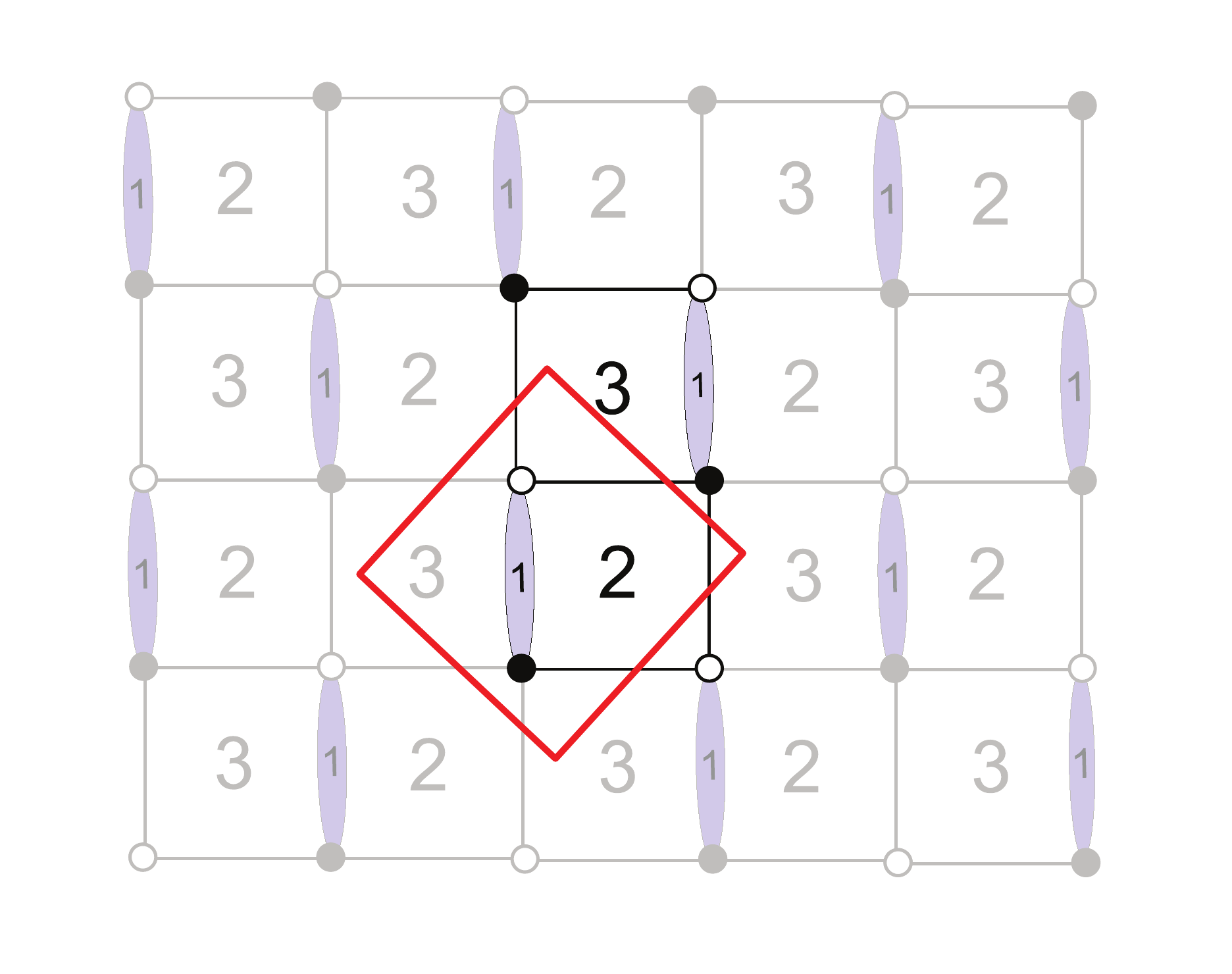}
\caption{Phase I of $\CC \times \BC$ with CS levels $k_1 =  -k_2 = 1$, $k_3 = 0$}
\label{ph1conxc}
\end{center}
\end{figure}
\begin{figure}[htbp]
\begin{center}
  \vskip 0.3cm
  \hskip -7cm
  \includegraphics[totalheight=0.9cm]{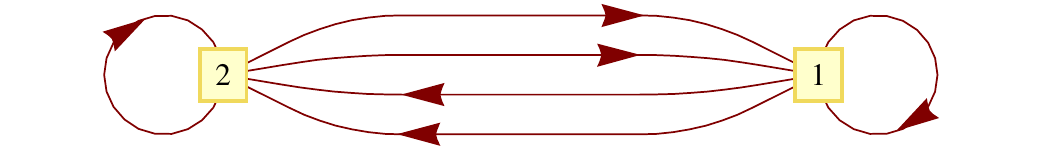}
  \vskip -1.4cm
  \hskip 4cm
  \includegraphics[totalheight=3.3cm]{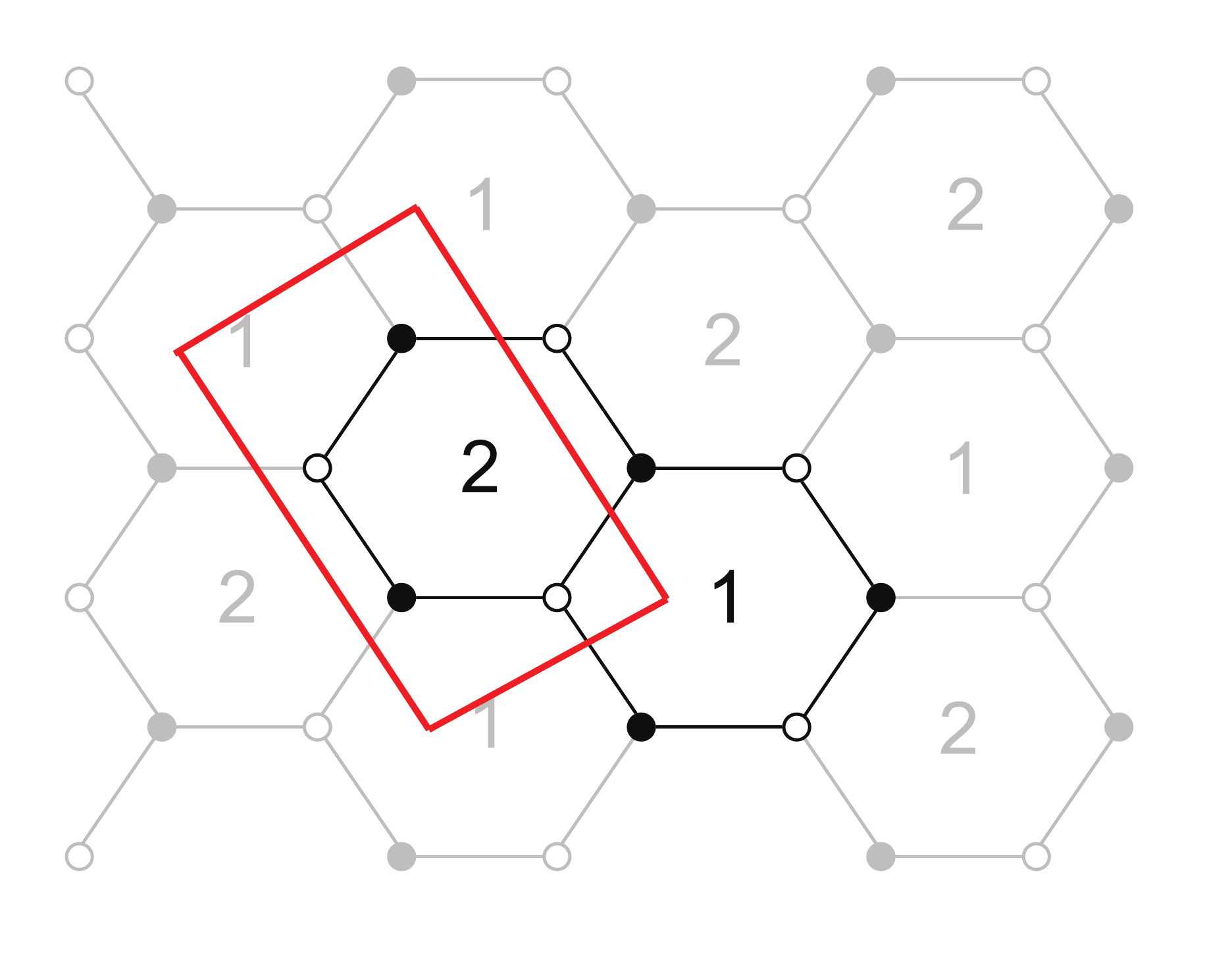}
  \caption{Phase II of $\CC \times \BC$ with CS levels $k_1 =  -k_2 = 1$}
\label{ph2conxc}
\end{center}
\end{figure}
\begin{figure}[htbp]
\begin{center}
  \vskip 0cm
  \hskip -7cm
  \includegraphics[totalheight=1cm]{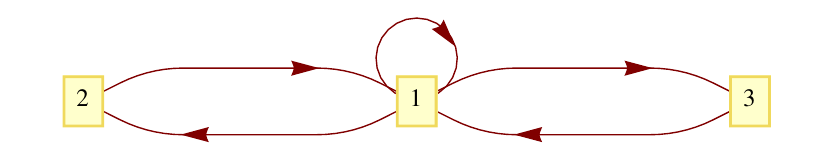}
  \vskip -1.5cm
  \hskip 3cm
  \includegraphics[totalheight=3.2cm]{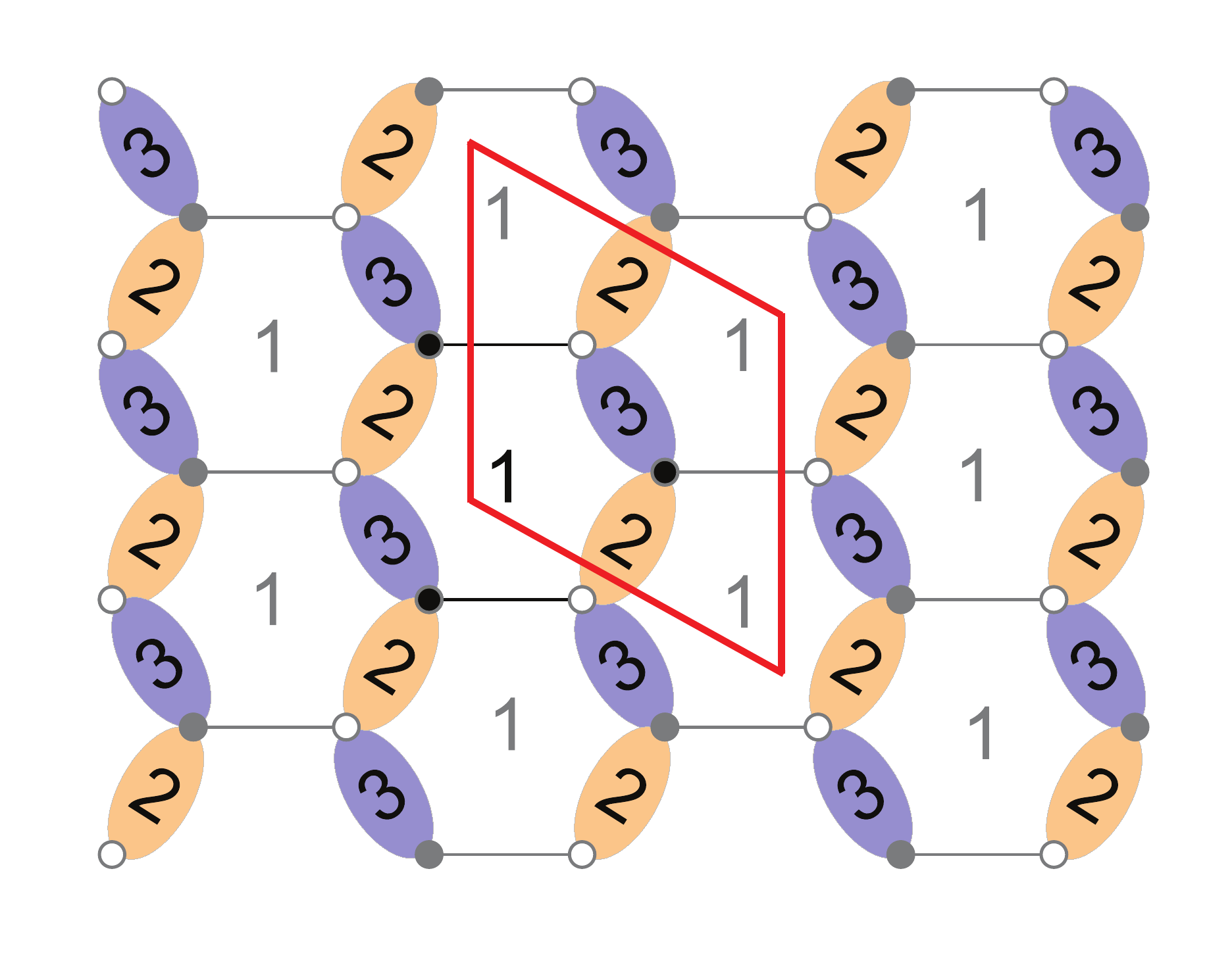}
    \caption{Phase III of $\CC \times \BC$ with CS levels $k_1 = 0$, $k_2 = -k_3 = 1$}
\label{ph3conxc}
\end{center}
\end{figure}

\begin{figure}[h]
\begin{center}
  \includegraphics[totalheight=2.5cm]{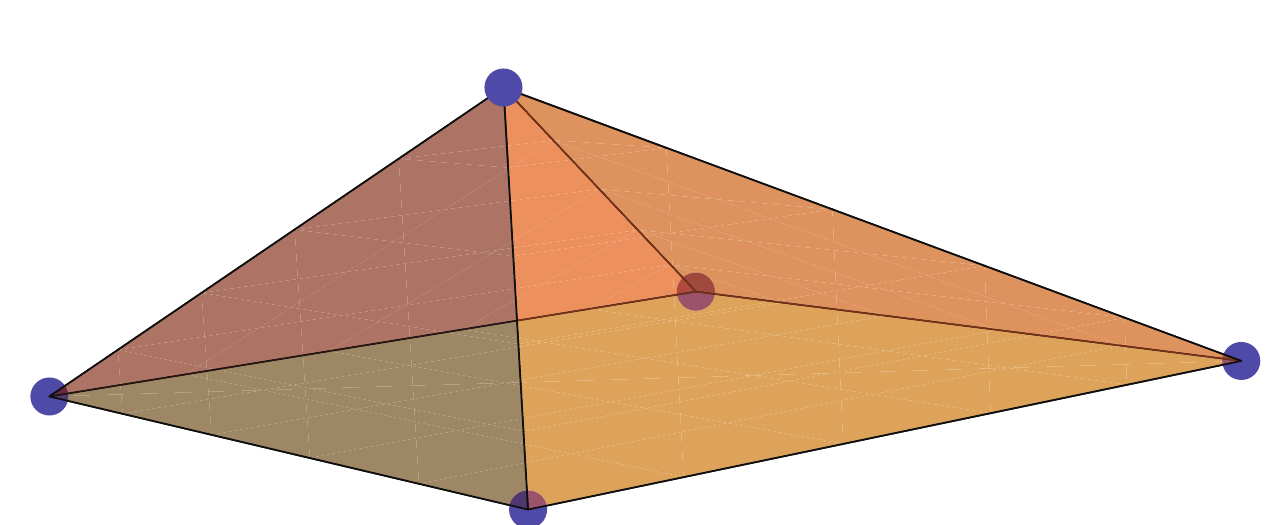}
 \caption{The toric diagram of the $\CC \times \BC$ theory. The 4 points at the corners on the base form the toric diagram of the conifold ($\CC$), and the point at the tip of the pyramid forms the toric diagram of $\BC$.}
  \label{f:torconxc}
\end{center}
\end{figure}

\subsubsection{A closer look at Phase II of $\CC \times \BC$} 
Let us focus on just one phase of the $\CC \times \BC$ theory. We summarise some of the interesting features of the model below:
\begin{itemize}
\item In $(3+1)$ dimensions, the quiver and tiling correspond to the $\BC^2/\BZ_2 \times \BC$ theory (\fref{ph2conxc}).
\item Since the number of gauge groups is $G=2$, it follows from \eref{mesonic} that the Master space is the same as the mesonic moduli space.
\item  From the superpotential
\bea
W = \tr \left( \phi_1 (X_{12}^2 X_{21}^1 - X_{12}^1 X_{21}^2 ) + {\phi}_2 (X_{21}^2 X_{12}^1 - X_{21}^1 X_{12}^2) \right)~,
\eea
it can be shown \cite{master} that the Master space (and hence the mesonic moduli space) is $\CC \times \BC$, where the conifold $\CC$ is parametrised by $X^1_{12}, X^2_{12}, X^1_{21}, X^2_{21}$ 
with the constraint $X^1_{12} X^2_{21} = X^2_{12} X^1_{21}$, and the $\BC$ is parametrised by $\phi_1 = \phi_2$.
\item It follows that $\phi_1, \phi_2$ are free fields, and so each of them has an R-charge of $1/2$. 
By symmetry, it can be seen that the requirement of R-charge 2 to $W$ divides $3/2$ equally among two $X$ fields.  Hence, each of the $X$ fields carries an R-charge of $3/4$.
\item Chiral fields have non-trivial scaling dimensions.  Hence, the IR fixed point is non-trivial.
\item The R-charges derived above agree with the computation by minimising the volume of the corresponding SE manifold \cite{Hanany:2008fj, phase}.  This provides a (weak) test of the AdS/CFT correspondence.   
\end{itemize}

\subsubsection{The global symmetry and charges}
The global symmetry of the $\CC \times \BC$ theory is $SU(2) \times SU(2) \times U(1)_q \times U(1)_R \times U(1)_B$.  The charges of the perfect matchings under the global symmetry are given in \tref{t:chargeconxc}.
The mesonic generators of each phase are listed in \tref{t:comgenconxc}.
The mesonic Hilbert series of $\CC \times \BC$ is 
\bea
\gm_1(t_1,t_2,x_1,x_2) 
&=& \frac{1}{1-t_2} \times  \frac{1-t_1^4}{\left(1-t^2_1 x_1 x_2\right)\left(1-\frac{t^2_1 x_2}{x_1}\right)\left(1-\frac{t^2_1 x_1}{x_2}\right)\left(1-\frac{t^2_1}{x_1 x_2}\right)} \nn \\
&=& \sum^{\infty}_{i=0} t_2^{i} \sum^{\infty}_{n=0}[n;n]t_1^{2n}~, \label{meshsd1c}
\eea
where $t_1 = t^3 q$ and $t_2 = t^4/q^4$.  Note that the first factor is the Hilbert series of $\BC$ and the second factor is the Hilbert series of $\CC$.
\begin{table}[h!]
 \begin{center} 
  \begin{tabular}{|c||c|c|c|c|c|c|}
  \hline
  \;& $SU(2)_1$&$SU(2)_2$&$U(1)_q$&$U(1)_B$&$U(1)_R$&  fugacity \\
  \hline \hline
  $p_1$&$1$&0&1&1&$3/8$ & $t^3 q b x_1$ \\
  \hline
  $p_2$&$-1$&0&1&1&$3/8$ & $t^3 q b/x_1$ \\
  \hline
  $p_3$&0&$1$&1&$-1$ &$3/8$ & $t^3 q x_2/b$\\
  \hline
  $p_4$&0&$-1$&1&$-1$&$3/8$ & $t^3 q/ (b x_2)$\\
  \hline
  $p_5$&0&0&$-4$&0&$1/2$ & $t^4/q^4$\\
  \hline
  \end{tabular}
  \end{center}
\caption{The global symmetry of the $\CC \times \BC$ theory. Here $t$ is the chemical potential (or strictly speaking the fugacity) associated with the $U(1)_{R}$ charges.  The power of $t$ counts R-charges in units of $1/8$, $q$ is the fugacity associated with the $U(1)_{q}$ charges, and $x_1,~x_2$ are respectively the $SU(2)_1,~SU(2)_2$ weights.}
\label{t:chargeconxc}
\end{table}
\begin{table}[h!]
\begin{center}
  \begin{tabular}{|c||c|c|c|}
    \hline
    Perfect & Generators  & Generators  & Generators  \\
 Matchings &of Phase I& of Phase II & of Phase III\\
    \hline
     $p_1 p_3$ & $X_{13}X^{1}_{32}$ & $X^{1}_{12}$ & $X_{21}X_{12}$ \\
     $p_2 p_3$ &  $X_{13}X^{2}_{32}$ & $X^{1}_{21}$  &$X_{21}X_{13}$ \\
     $p_1 p_4$ & $X_{23}X^{1}_{32}$ & $X^{2}_{12}$ &$X_{31}X_{12}$  \\
     $p_2 p_4$ & $X_{23}X^{2}_{32}$ & $X^{2}_{21}$  &$X_{21}X_{13}$ \\
     $p_5$ & $X_{21}$ & $\phi_1=\phi_2$& $\phi_{1}$ \\
     \hline
  \end{tabular}
  \end{center}
 \caption{A comparison between the generators of different phases of the $\CC \times \BC$ theory.  In terms of the perfect matchings, the generators of different phases are precisely the same.}
 %In Phase I, we require gauge invariance with respect to the gauge group 3, and so the indices corresponding to the gauge group 3 are contracted.  In Phase III, we require gauge invariance with respect to the gauge group 1, and so the indices corresponding to the gauge group 1 are contracted.}
\label{t:comgenconxc}
\end{table}

\section{M2-brane theories and Fano 3-folds}
In this section, we focus on gauge theories arising from M2-branes probing CY 4-fold singularities that can be realised as cones over smooth toric Fano 3-folds. These Fano varieties have already attracted much mathematical interest and a complete classification of these geometries is known \cite{fano3}.  There are precisely 18 smooth toric Fano 3-folds \cite{toricfano3, database}.  In this paper, we present the gauge theories corresponding to 5 of them, namely $\CP^2 \times \CP^1$, $\CP^1 \times \CP^1 \times \CP^1$, $dP_n \times \CP^1$ ($n=1,2,3$). For more information about the others, we refer the reader to \cite{Hanany:2009vx} and the work in progress \cite{future}.

\subsection{The $M^{1,1,1}$ Theory}

The quiver and tiling are given in \fref{f:m111}.  In $(3+1)$ dimensions, this corresponds to the $dP_0$ theory.  Let us assign the CS levels $\vec{k} = (1, -2, 1)$. The superpotential is $W= \tr \left( \epsilon_{ijk} X^i_{12} X^j_{23} X^k_{31} \right)$.

\begin{figure}[ht]
 \centerline{  \epsfxsize = 3cm \epsfbox{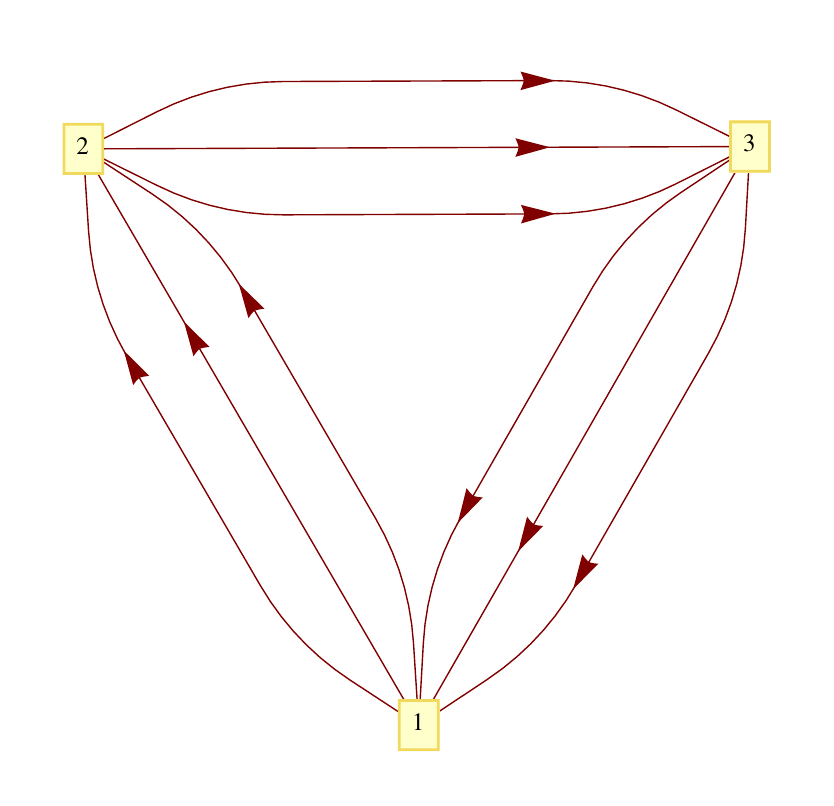}  \hskip 10mm \epsfxsize = 4cm \epsfbox{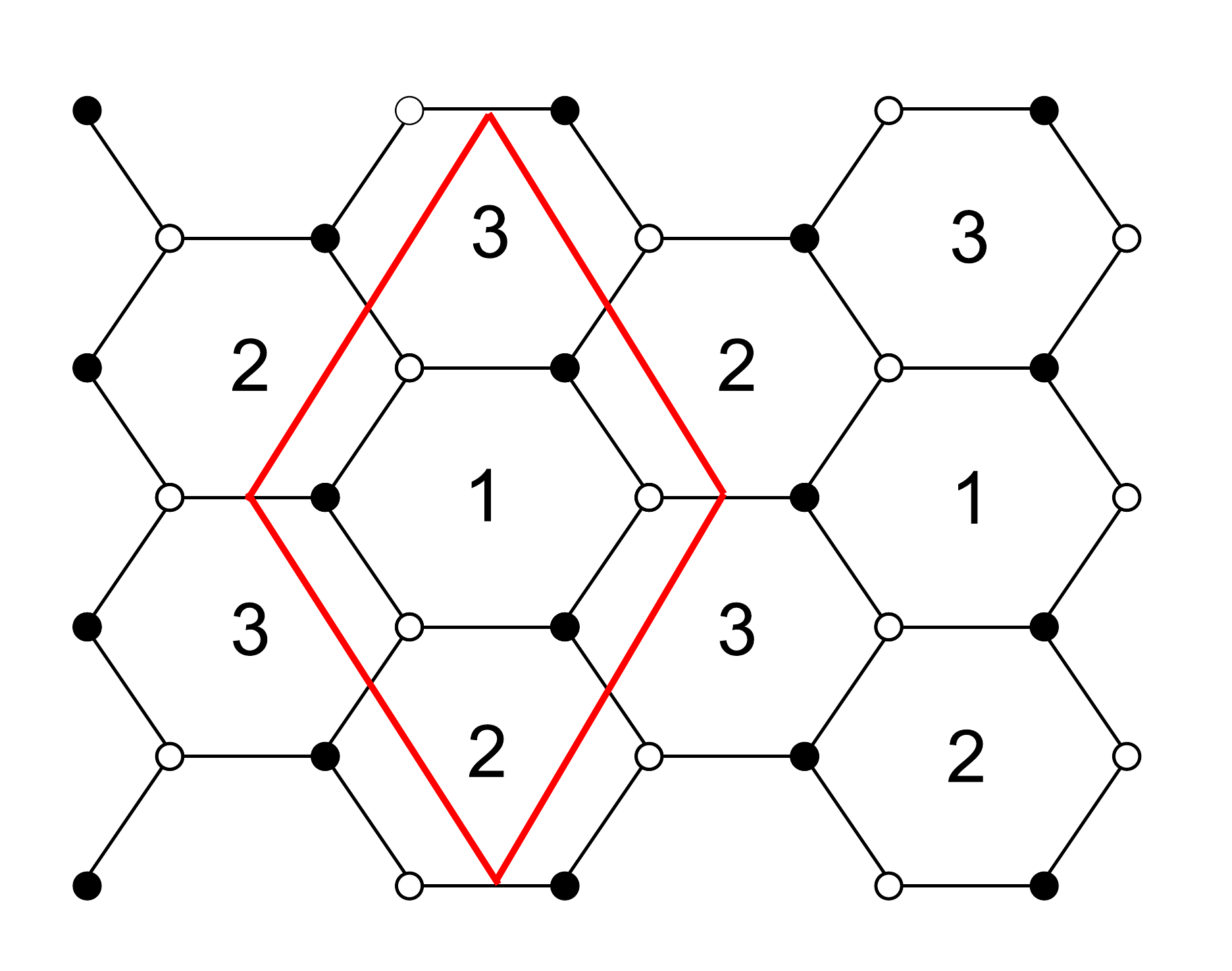}}
\caption{(i) Quiver diagram of the $M^{1,1,1}$  theory.\ (ii) Tiling of the $M^{1,1,1}$  theory.}
  \label{f:m111}
\end{figure}
\begin{figure}[ht]
\begin{center}
  \includegraphics[totalheight=2cm]{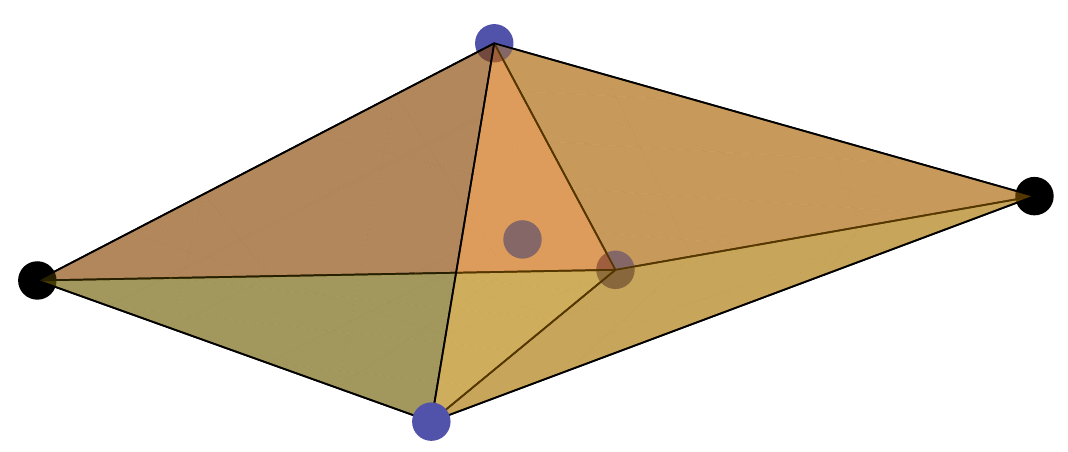}
 \caption{The toric diagram of the $M^{1,1,1}$ theory.}
  \label{f:tdtoricfano24}
\end{center}
\end{figure}

The toric diagram of this theory is given by \fref{f:tdtoricfano24}.
Note that the 4 blue points form the toric diagram of $\CP^2$ , and the 2 black points together with the blue internal point form the toric diagram of $\CP^1$.
Hence, this theory corresponds to the cone over $\CP^2 \times \CP^1$.

The mesonic global symmetry of this theory is $SU(3) \times SU(2) \times U(1)_R$. There is also one baryonic $U(1)_B$ symmetry. The charges of the perfect matchings under these symmetries are listed in
\tref{t:chargefano24}.  The Hilbert series of the mesonic moduli space is given by
\bea
\gm (t,x,y_1,y_2; M^{1,1,1}) = \sum^{\infty}_{n=0}\left[3n,0;2n\right]t^{18n}~.
\eea
This is a sum over all irreducible representations of the form $[3n, 0; 2n]$, where the first two numbers are highest weights of an $SU(3)$ representation (totally symmetric $3n$ tensor), and the last number is the highest weight of an $SU(2)$ representation (of spin $n$). Indeed, this result confirms the known KK spectrum on $M^{1,1,1}$ \cite{Fabbri:1999hw}.
\begin{table}[h]
 \begin{center}  % put inside center environment
  \begin{tabular}{|c||c|c|c|c|c|c|}
  \hline
  \;& $SU(3)$&$SU(2)$&$U(1)_R$&$U(1)_B$&fugacity\\
  \hline  \hline 
  $p_1$&$(1,0)$&$0$&4/9&0&$t^4 y_1  $\\
  \hline
  $p_2$&$(-1,1)$&$0$&4/9&0&$t^4 y_2 / y_1 $\\
  \hline
  $p_3$&$(0,-1)$&$0$&4/9&0&$t^4 / y_2 $\\
  \hline
  $r_1$&(0,0)&1&1/3&$-1$&$t^3 x / b$ \\
  \hline
  $r_2$&(0,0)&$-1$&1/3&$-1$&$t^3/(x b)$\\
  \hline
  $s_1$&(0,0)&0&0&2&$  b^2$\\
  \hline
  \end{tabular}
  \end{center}
\caption{Charges of the perfect matchings under the global symmetry of the $M^{1,1,1}$ theory. Here $t$ is the fugacity of the R-charge (in multiples of $1/9$), $y_1,y_2$ are the fugacities of the $SU(3)$ symmetry, $x$ is the fugacity of the $SU(2)$ symmetry and $b$ is the fugacity of the $U(1)_B$ symmetry.  We have used the notation $(a,b)$ to represent a weight of $SU(3)$.}
\label{t:chargefano24}
\end{table}
 
\subsection{The $Q^{1,1,1}/\BZ_2$ Theory}
There are two known toric phases of this theory. Their quivers and tilings are given in Figures \ref{f:phase1f0} and \ref{f:phase2f0}. The toric digram is drawn in \fref{f:torq111z2}.
This theory corresponds to the cone over $\BP^1 \times \BP^1 \times \BP^1$.

\begin{figure}[h!]
\begin{center}
  \vskip -0.5cm
  \hskip -7cm
  \includegraphics[totalheight=3.0cm]{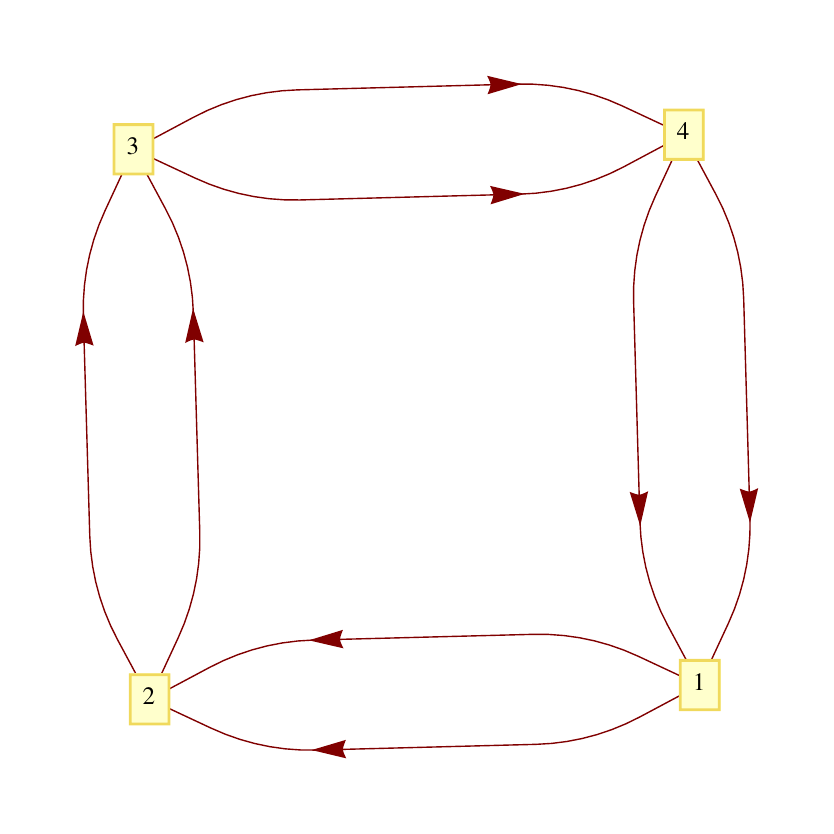}
    \vskip -2.8cm
  \hskip 6cm
  \includegraphics[totalheight=2.5cm]{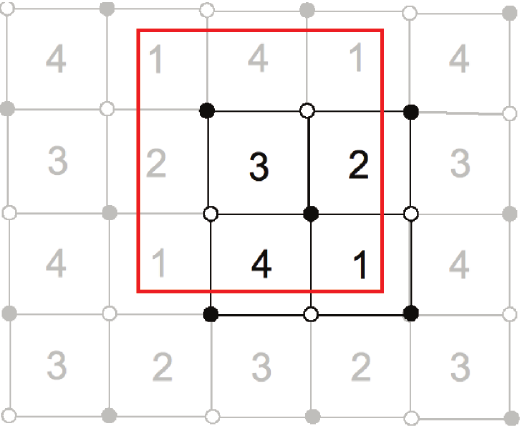}
 \caption{The quiver and tiling of Phase I of $Q^{1,1,1}/\BZ_2$ with $\vec{k}=(1,-1,-1,1)$.  The superpotential is $W = \epsilon_{ij} \epsilon_{pq} \tr(X_{12}^i X_{23}^p X_{34}^j X_{41}^q)$.}
  \label{f:phase1f0}
\end{center}
\end{figure}
\begin{figure}[h!]
\begin{center}
  \vskip 1cm
  \hskip -7cm
  \includegraphics[totalheight=2cm]{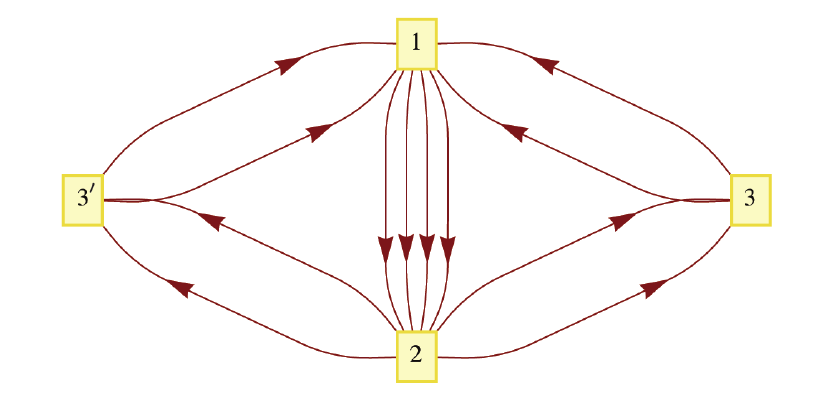}
    \vskip -3.0cm
  \hskip 6cm
  \includegraphics[totalheight=2.5cm]{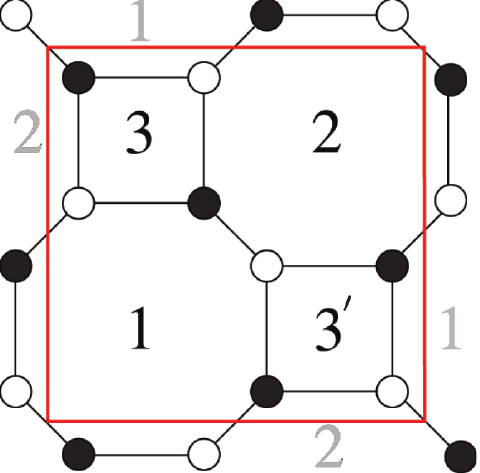}
 \caption{{\footnotesize The quiver and tiling of Phase II of $Q^{1,1,1}/\BZ_2$ with $k_1=k_2=-k_3=-k_{3'}=1$. The superpotential is $W = \epsilon_{ij}\epsilon_{kl} \tr(X^{ik}_{12}X^{l}_{23} X^{j}_{31}) - \epsilon_{ij}\epsilon_{kl} \tr(X^{ki}_{12}X^{l}_{23'}X^{j}_{3'1})$.}}
  \label{f:phase2f0}
\end{center}
\end{figure}
\begin{figure}[h!]
\begin{center}
  \includegraphics[totalheight=4cm]{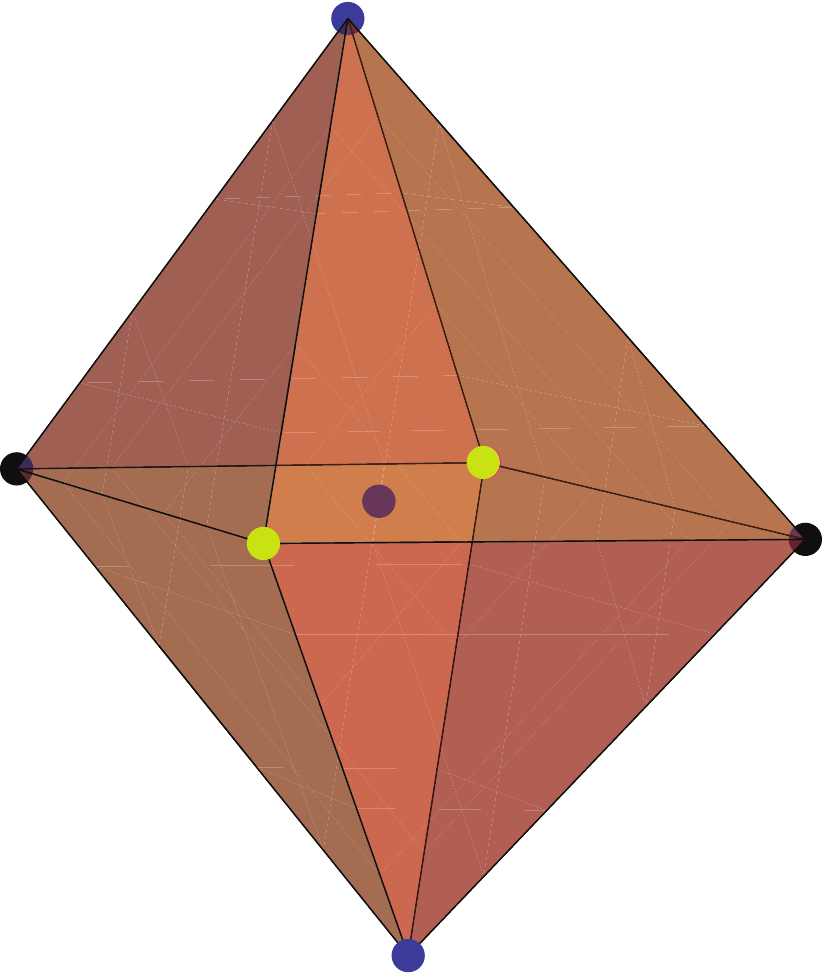}
 \caption{The toric diagram of $Q^{1,1,1}/\BZ_2$.  Note that the 3 blue points form the toric diagram of $\BP^1$, and so as the yellow points (together with the internal points) and the black points (together with the internal point). Thus, this theory corresponds to the cone over $\BP^1 \times \BP^1 \times \BP^1$.}
  \label{f:torq111z2}
\end{center}
\end{figure}

The mesonic symmetry of this model is $SU(2)^3 \times U(1)_R$.   There are two baryonic charges.  The charges of perfect matchings under these symmetries are given in \tref{chargeph1f0}.
The Hilbert series of the mesonic moduli space can be written as
\bea
\gm_1 (t,x_1,x_2,x_3; Q^{1,1,1}/\BZ_2) 
= \sum^{\infty}_{n=0}[2n;2n;2n]t^{6n}~.
\label{meshsph1fo}
\eea

\begin{table}[h!]
 \begin{center}  % put inside center environment
  \begin{tabular}{|c||c|c|c|c|c|c|c|}
  \hline
  \;& $SU(2)_{1}$&$SU(2)_{2}$&$SU(2)_{3}$&$U(1)_R$&$U(1)_{B_1}$&$U(1)_{B_2}$& fugacity\\
  \hline \hline  % put a line under headers
  $p_1$& $1$&$0$&$0$&$1/3$&$1$&$0$& $t b_1 x_1$ \\
  $p_2$& $-1$&$0$&$0$&$1/3$&$1$&$0$ & $t b_1/x_1$ \\
  $q_1$& $0$&$1$&$0$&$1/3$&$0$&$0$  & $t x_2$\\
  $q_2$& $0$&$-1$&$0$&$1/3$&$0$&$0$& $ t / x_2$\\
  $r_1$& $0$&$0$&$1$&$1/3$&$-1$&$-1$& $ t x_3/(b_1 b_2)$\\
  $r_2$& $0$&$0$&$-1$&$1/3$&$-1$&$-1$& $ t / (x_3  b_1 b_2)$\\
  $s_1$& $0$&$0$&$0$&$0$&$0$&$2$&  $b_2^2$ \\
  $s_2$& $0$&$0$&$0$&$0$&$0$&$0$ & $1$ \\
  $s_3$& $0$&$0$&$0$&$0$&$0$&$0$ & $1$ \\ 
  \hline
  \end{tabular}
  \end{center}
  \caption{The global symmetry of the $Q^{1,1,1}/\BZ_2$ theory. Here $t$ is the fugacity of R-charge, $x_1,x_2,x_3$ are weights of $SU(2)_{1}, SU(2)_{2}, SU(2)_3$, and $b_1, b_2$ are baryonic fugacities of $U(1)_{B_1}, U(1)_{B_2}$. Note that the perfect matching $s_3$ does not exist in Phase I but exists in Phase II.}
\label{chargeph1f0}
\end{table}

\subsection{The $dP_n \times \CP^1$ Theories} 
Tilings have been found that correspond to the cones over $dP_n \times \CP^1$, for $1 \leq n \leq 3$. 
We present both the quiver diagrams and tilings Figures \ref{f:tqfano123}, \ref{f:fano266tileandquiver}, \ref{f:fano324tileandquiver} and their corresponding toric data \fref{f:toricdpnxp1}. Full details of these models will be presented in future work \cite{future}.
\begin{figure}[h!]
\begin{center}
 \centerline{  \epsfxsize = 3cm \epsfbox{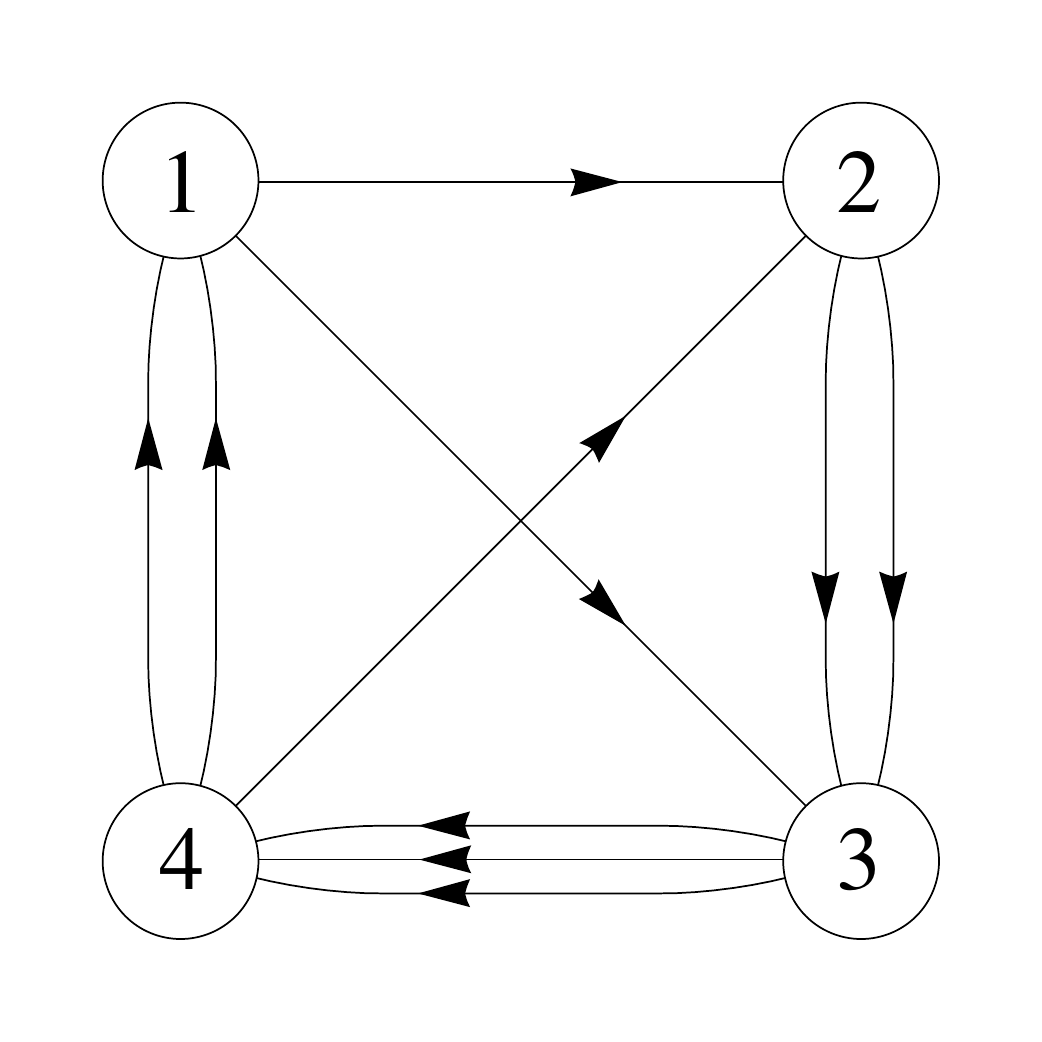}\hskip 10mm \epsfxsize = 4cm \epsfbox{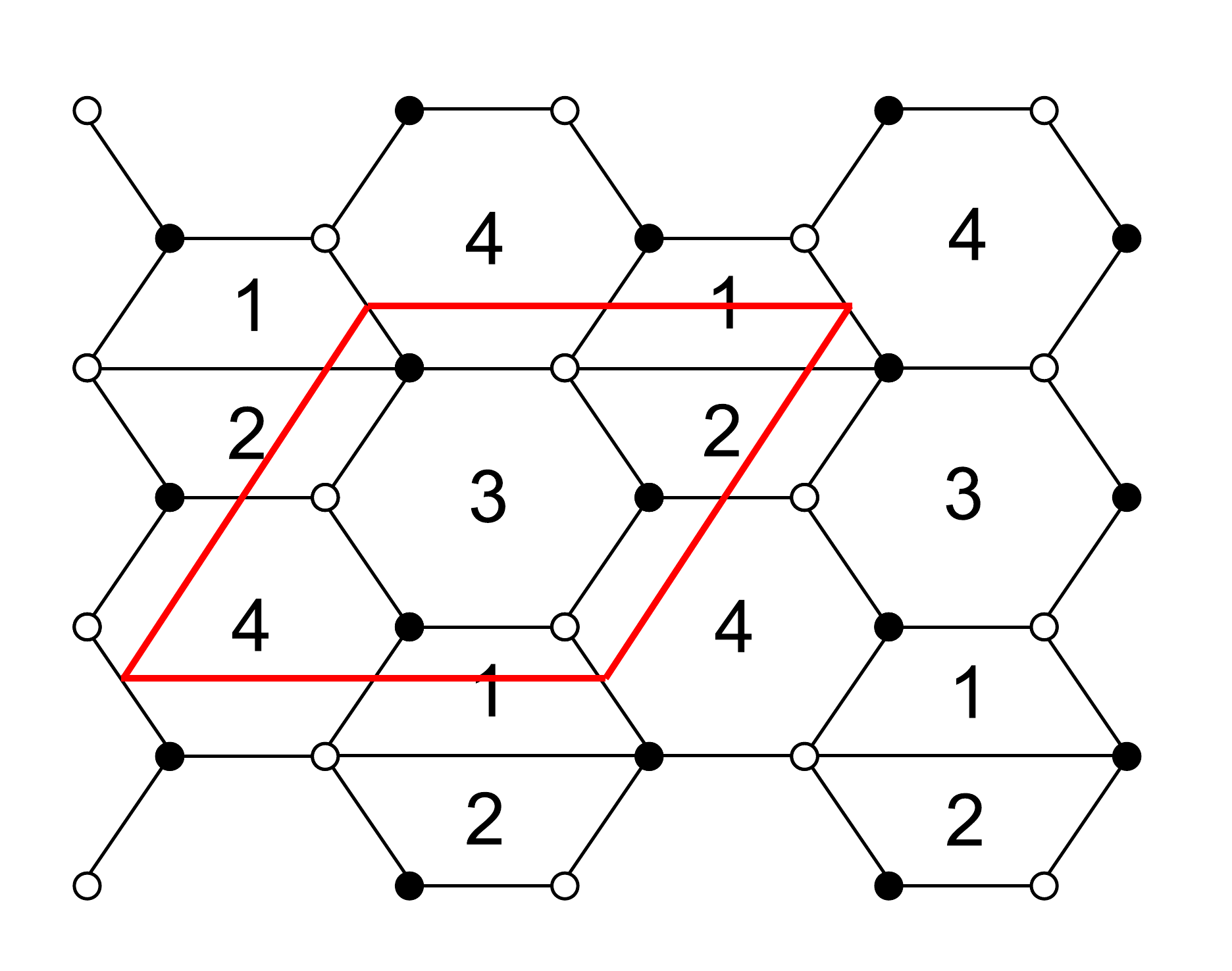}}
 \caption{{\footnotesize [The $dP_1 \times \BP^1$ theory] (i) Quiver diagram \ (ii) Tiling. The Chern-Simons levels are $\vec{k} = (1,1,-1,-1)$. The superpotential is 
 $W = \tr \left[ \epsilon_{ij} \left( X_{13} X^{i}_{34} X^{j}_{41} + X_{42} X^{i}_{23} X^{j}_{34} + X_{12} X^{j}_{23} X^{3}_{34} X^{i}_{41}\right) \right]$.}}
  \label{f:tqfano123}
\end{center}
\end{figure}
\begin{figure}[h!]
 \centerline{  \epsfxsize = 3cm \epsfbox{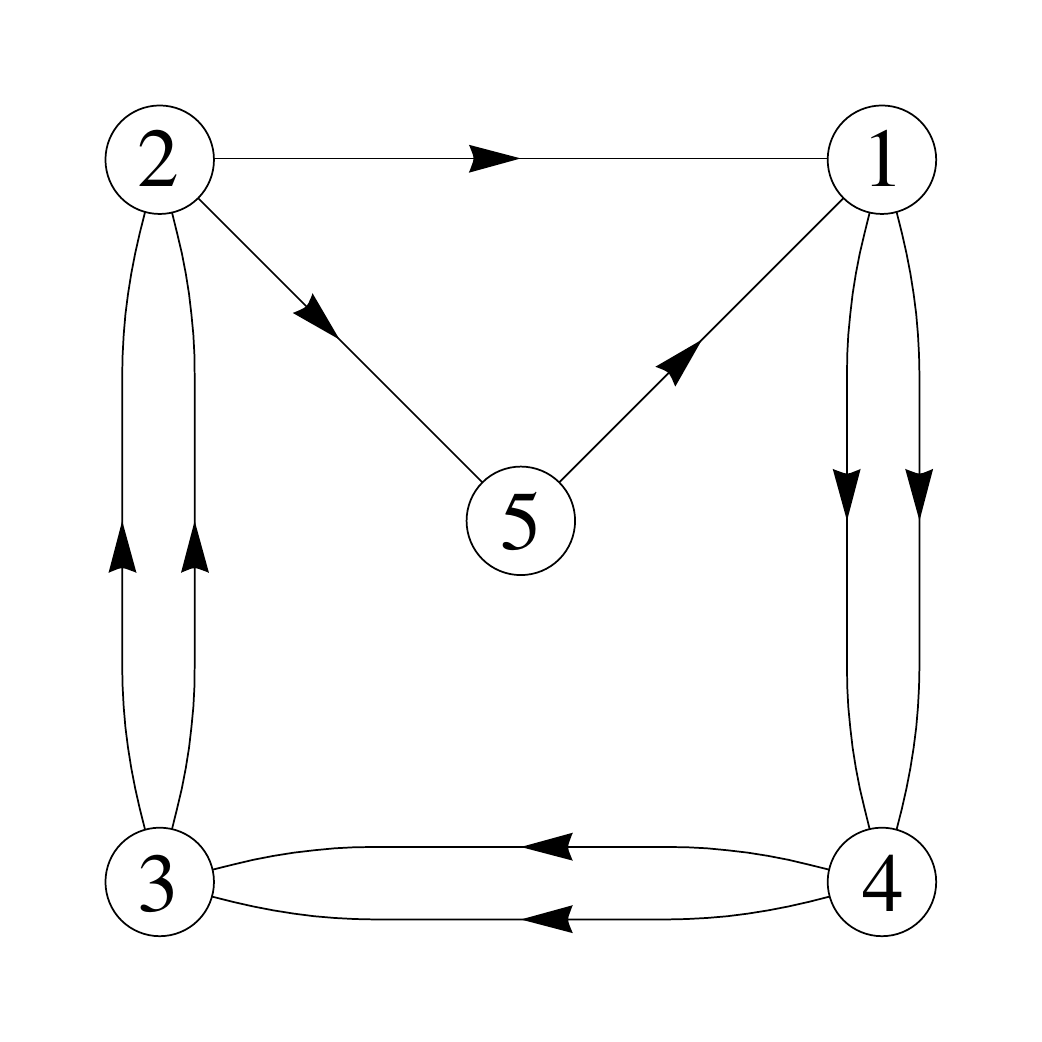}\hskip 8mm \epsfxsize = 4cm \epsfbox{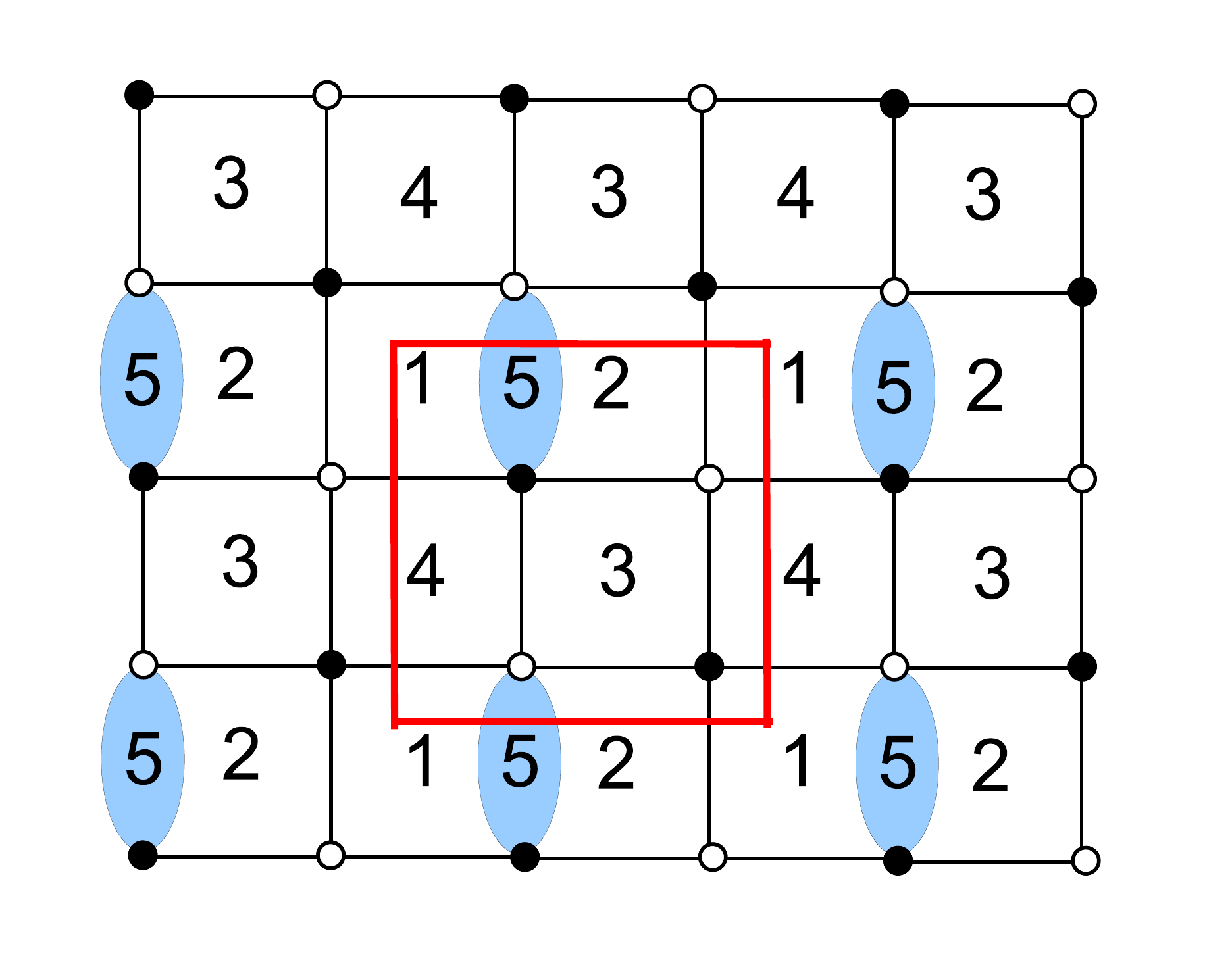}}
 \caption{{\footnotesize [The $dP_2 \times \BP^1$ theory] (i) Quiver diagram \ (ii) Tiling. The Chern-Simons levels are $\vec{k} = (1, 1, 
-1, 0, -1)$. The superpotential is 
$W = \tr \left[ \epsilon_{ij} ( X_{43}^1 X_{32}^j X_{21} X_{14}^i + X_{51} X_{14}^j X_{43}^2 X_{32}^i X_{25} ) \right]$.}}
  \label{f:fano266tileandquiver}
\end{figure} 
\begin{figure}[h!]
 \centerline{  \epsfxsize = 3cm \epsfbox{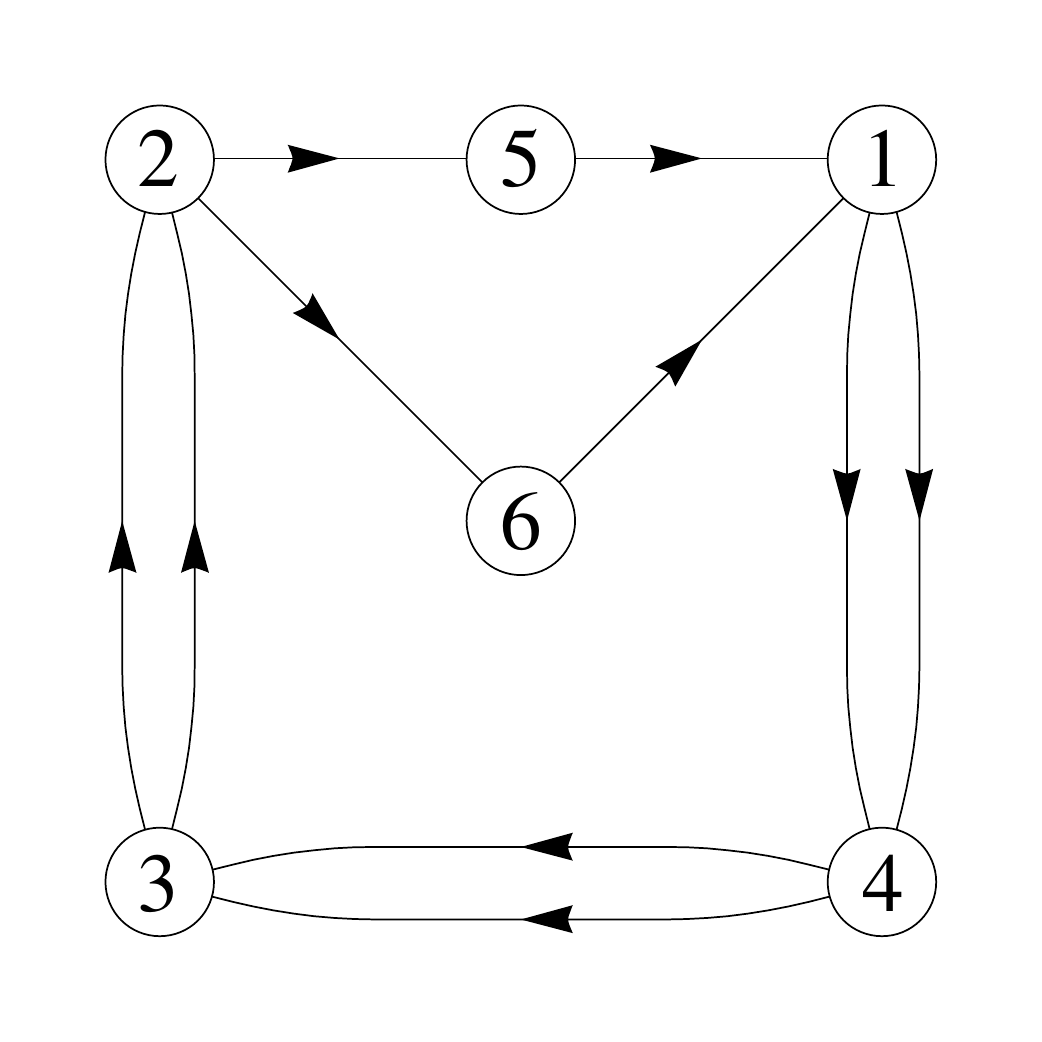}\hskip 8mm \epsfxsize = 4cm \epsfbox{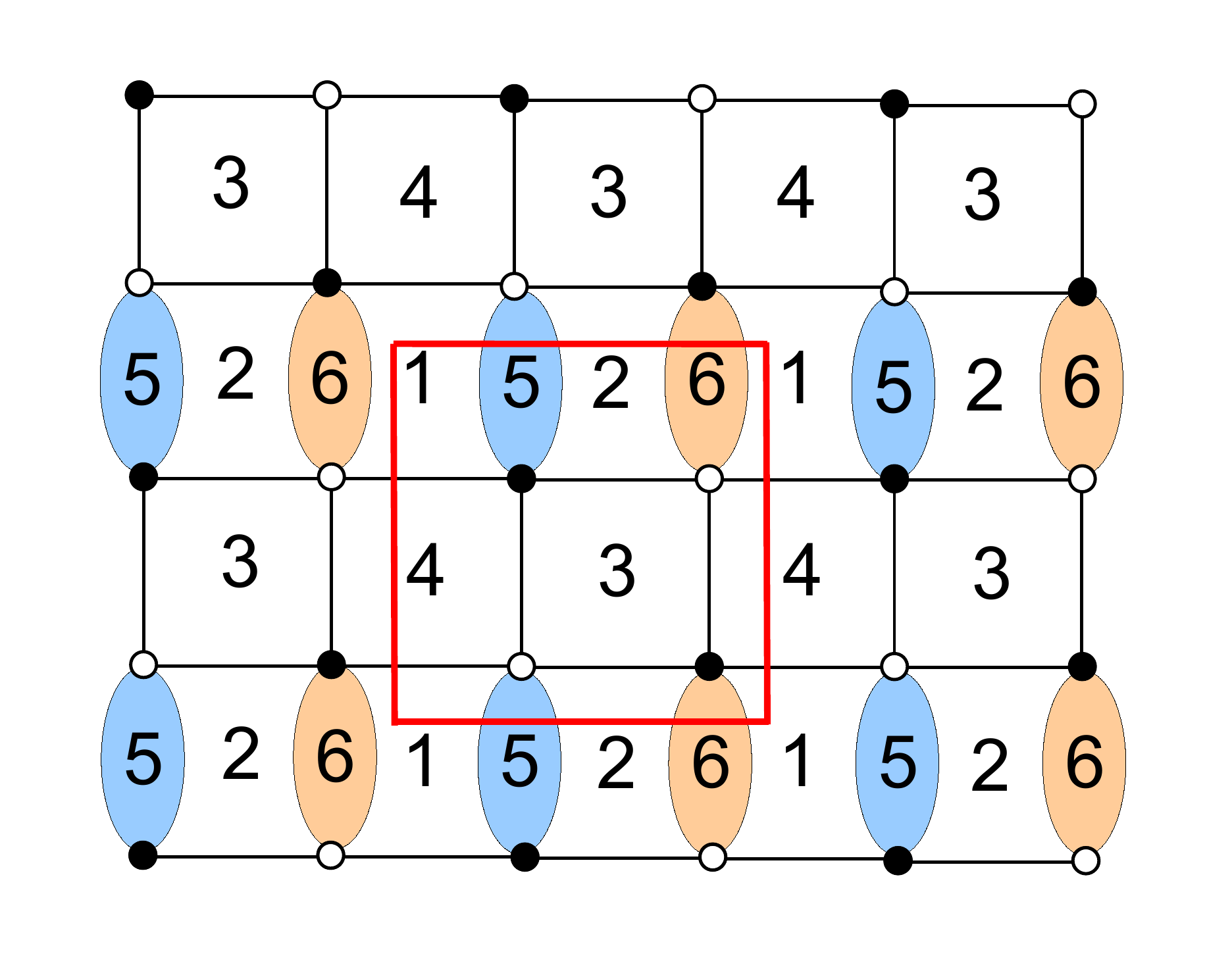}}
%\vskip -1cm
%\hskip 2cm
%\includegraphics[totalheight=3cm]{QuiverFano324.pdf}
%\vskip 0.1cm
%\hskip 0.7cm
%\includegraphics[totalheight=3cm]{fano324tilinggg.pdf}
 \caption{{\footnotesize [The $dP_3 \times \BP^1$ theory] (i) Quiver diagram \ (ii) Tiling. The Chern-Simons levels are $\vec{k}=(0,-1,0,-1,1,1)$. The superpotential is 
$W = \tr \left[\epsilon_{ij} \left( X_{14}^j X_{43}^2 X_{32}^i X_{26} X_{61} + X_{14}^i X_{43}^1 X_{32}^j X_{25} X_{51} \right) \right]$.}}
  \label{f:fano324tileandquiver}
\end{figure}
 
\begin{figure}[h!]
%\vskip -1cm
\hskip -2cm
\includegraphics[totalheight=2.5cm]{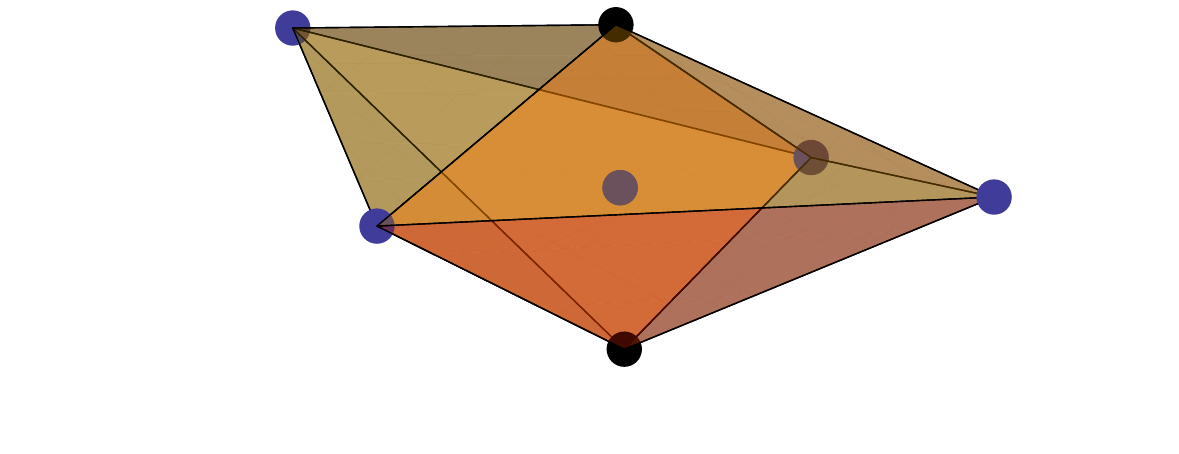}
\hskip -1.5cm
\includegraphics[totalheight=2.5cm]{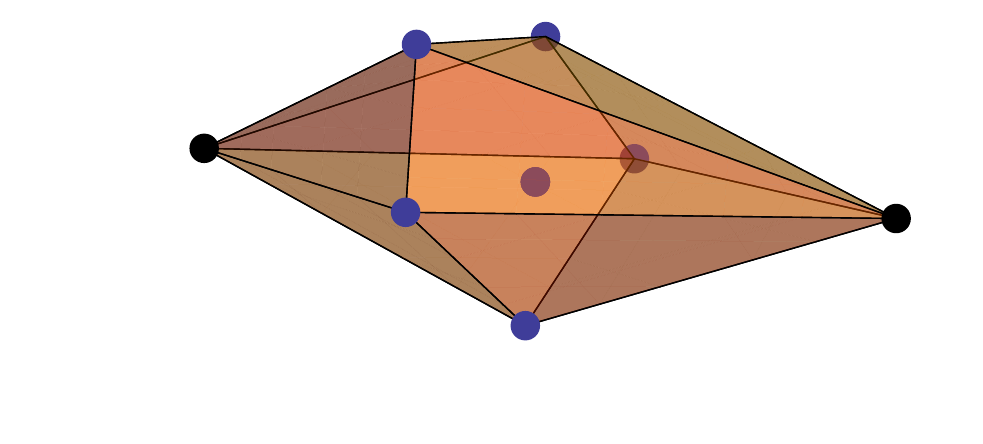}
%\vskip -0.2cm
\hskip -1cm
\includegraphics[totalheight=2cm]{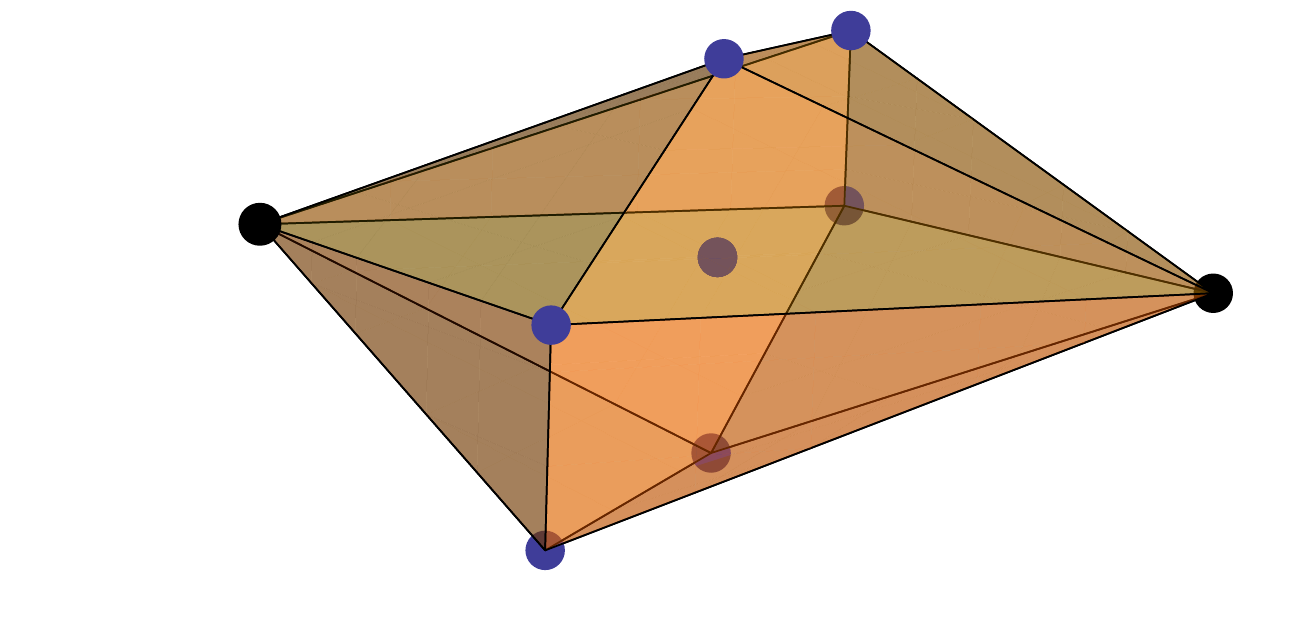}
 \caption{{\footnotesize The toric diagrams of (i) left: the $dP_1 \times \CP^1$ theory, (ii) middle: the $dP_2 \times \CP^1$ theory,  (ii) right: the $dP_3 \times \CP^1$ theory. In each figure, the blue points form the toric diagram of $dP_n$ and the black points (together with the blue internal point) form a toric diagram of $\CP^1$.}}
  \label{f:toricdpnxp1}
\end{figure}


\newpage
 \begin{thebibliography}{99}

\footnotesize{

%1
\bibitem{HullEtAl}
  B.~S.~Acharya, J.~M.~Figueroa-O'Farrill, C.~M.~Hull and B.~J.~Spence,
%  ``Branes at conical singularities and holography,''
  Adv.\ Theor.\ Math.\ Phys.\  {\bf 2} (1999) 1249
  [arXiv:hep-th/9808014].
  %%CITATION = 00203,2,1249;%%
%\cite{Hanany:2008cd}
%2
\bibitem{Plesser}
  D.~R.~Morrison and M.~R.~Plesser,
% ``Non-spherical horizons. I,''
  Adv.\ Theor.\ Math.\ Phys.\  {\bf 3} (1999) 1
  [arXiv:hep-th/9810201].
%\cite{Aharony:2008ug}
%\cite{Ueda:2008hx}
%3
 \bibitem{Martelli:2008si}
  D.~Martelli and J.~Sparks,
 % ``Moduli spaces of Chern-Simons quiver gauge theories and AdS(4)/CFT(3),''
  arXiv:0808.0912 [hep-th].
  %4  
\bibitem{Hanany:2008cd}
  A.~Hanany and A.~Zaffaroni,
  %``Tilings, Chern-Simons Theories and M2 Branes,''
  arXiv:0808.1244 [hep-th].
%5
 \bibitem{Ueda:2008hx}
  K.~Ueda and M.~Yamazaki,
 % ``Toric Calabi-Yau four-folds dual to Chern-Simons-matter theories,''
  arXiv:0808.3768 [hep-th].
   %%%%%%%%%%%%%%%%%%%%%%%%%%%%%%%%%%%%%%%%%%%%%%%%% quiver
%6  
  \bibitem{taxonomy}
  A.~Hanany and Y.~H.~He,
%  ``M2-Branes and Quiver Chern-Simons: A Taxonomic Study,''
  arXiv:0811.4044 [hep-th].
  %%CITATION = ARXIV:0811.4044;%% 
  %7
  %\cite{Hewlett:2009bx}
\bibitem{Hewlett:2009bx}
  J.~Hewlett and Y.~H.~He,
  %``Probing the Space of Toric Quiver Theories,''
  arXiv:0909.2879 [hep-th].
  %%CITATION = ARXIV:0909.2879;%%
  %8
%%%%%%%%%%%%%%%%%%%%%%%%%%%%%% tiling
\bibitem{Hanany:2008fj}
  A.~Hanany, D.~Vegh, A.~Zaffaroni,
 % ``Brane Tilings and M2 Branes,''
  arXiv:0809.1440.
  %9
  \bibitem{phase}
  J.~Davey, A.~Hanany, N.~Mekareeya and G.~Torri,
%  ``Phases of M2-brane Theories,''
  arXiv:0903.3234 [hep-th].
  %%CITATION = ARXIV:0903.3234;%%
  %10
  %\cite{Davey:2009qx}
\bibitem{higgs}
  J.~Davey, A.~Hanany, N.~Mekareeya and G.~Torri,
  %``Higgsing M2-brane Theories,''
  arXiv:0908.4033 [hep-th].
  %%CITATION = ARXIV:0908.4033;%%
  %11

%\cite{Benishti:2009ky}
\bibitem{unhiggs}
  N.~Benishti, Y.~H.~He and J.~Sparks,
  %``(Un)Higgsing the M2-brane,''
  arXiv:0909.4557 [hep-th].
  %%CITATION = ARXIV:0909.4557;%%


 %\cite{Davey:2009bp}
\bibitem{Davey:2009bp}
  J.~Davey, A.~Hanany and J.~Pasukonis,
  %``On the Classification of Brane Tilings,''
  arXiv:0909.2868 [hep-th].
  %%CITATION = ARXIV:0909.2868;%% 

  
 %%%%%%%%%%%%%%%%%%%%%%%%%%%%%%%%%% 3+1 tilings
%12
\bibitem{Hanany:2005ve}
  A.~Hanany and K.~D.~Kennaway,
  %``Dimer models and toric diagrams,'' 
  hep-th/0503149.
%13
\bibitem{Franco:2005rj}
  S.~Franco, A.~Hanany, K.~D.~Kennaway, D.~Vegh and B.~Wecht,
%  ``Brane dimers and quiver gauge theories,''
  JHEP {\bf 0601}, 096 (2006)
  [arXiv:hep-th/0504110].
  %14
%\cite{Hanany:2005ss}
\bibitem{Hanany:2005ss}
  A.~Hanany and D.~Vegh,
%  ``Quivers, tilings, branes and rhombi,''
  JHEP {\bf 0710}, 029 (2007)
  [arXiv:hep-th/0511063].
  %%CITATION = JHEPA,0710,029;%%
%15
\bibitem{Franco:2005sm}
  S.~Franco, A.~Hanany, D.~Martelli, J.~Sparks, D.~Vegh and B.~Wecht,
%  ``Gauge theories from toric geometry and brane tilings,''
  JHEP {\bf 0601}, 128 (2006)
  [arXiv:hep-th/0505211].
%16
%\bibitem{Feng:2005gw}
%  B.~Feng, Y.~H.~He, K.~D.~Kennaway and C.~Vafa,
 % ``Dimer models from mirror symmetry and quivering amoebae,''
%  Adv.\ Theor.\ Math.\ Phys.\  {\bf 12}, 3 (2008)
%  [arXiv:hep-th/0511287].
  %17
  %\cite{Broomhead:2008an}
%\bibitem{Broomhead:2008an}
%  N.~Broomhead,
%  ``Dimer models and Calabi-Yau algebras,''
%  arXiv:0901.4662 [math.AG].
%  %%CITATION = ARXIV:0901.4662;%% 
%%%%%%%%%%%%%%%%%%%%%%%%%%%%%%%%% Tiling reviews
 %18
 %\cite{Kennaway:2007tq}
\bibitem{Kennaway:2007tq}
  K.~D.~Kennaway,
 % ``Brane Tilings,''
  Int.\ J.\ Mod.\ Phys.\  A {\bf 22}, 2977 (2007)
  [arXiv:0706.1660 [hep-th]].
  %%CITATION = IMPAE,A22,2977;%%
  %19
  %\cite{Yamazaki:2008bt}
\bibitem{Yamazaki:2008bt}
  M.~Yamazaki,
 % ``Brane Tilings and Their Applications,''
  Fortsch.\ Phys.\  {\bf 56}, 555 (2008)
  [arXiv:0803.4474 [hep-th]].
  %%CITATION = FPYKA,56,555;%%
   %%%%%%%%%%%%%%%%%%%%%%%%%%%%%%%%%%%%%%%%%%%%%%%%%
%20
  %\cite{Amariti:2009rb}
\bibitem{Amariti:2009rb}
  A.~Amariti, D.~Forcella, L.~Girardello and A.~Mariotti,
%  ``3D Seiberg-like Dualities and M2 Branes,''
  arXiv:0903.3222 [hep-th].
  %%CITATION = ARXIV:0903.3222;%%
%21
     %\cite{Hanany:2009vx}
    \bibitem{Hanany:2009vx}
  A.~Hanany and Y.~H.~He,
  %``Chern-Simons: Fano and Calabi-Yau,''
  arXiv:0904.1847 [hep-th].
  %%CITATION = ARXIV:0904.1847;%% 
%22
\bibitem{toricfano3}
V.~V.~Batyrev, %``Toroidal Fano 3-folds'', 
Math. USSR-Izv. 19, 13-25 (1982)
%23
%\bibitem{toricfano4}
%V.~V.~Batyrev,% ``On the Classification of Toric Fano 4-folds'',
% arXiv:math/9801107.
%24
\bibitem{database}
\verb|http://malham.kent.ac.uk/grdb/index.php|  
%%%%%%%%%%%%%%%%%%%%%%%%%%%%%%%%%% random
%25
\bibitem{Feng:2000mi}
  B.~Feng, A.~Hanany and Y.~H.~He,
%  ``D-brane gauge theories from toric singularities and toric duality,''
  Nucl.\ Phys.\  B {\bf 595}, 165 (2001)
  [arXiv:hep-th/0003085].
  %26
  %\cite{Feng:2001xr}
\bibitem{Feng:2001xr}
  B.~Feng, A.~Hanany and Y.~H.~He,
  %``Phase structure of D-brane gauge theories and toric duality,''
  JHEP {\bf 0108}, 040 (2001)
  [arXiv:hep-th/0104259].
  %%CITATION = JHEPA,0108,040;%%


%\cite{Feng:2002fv}
\bibitem{Feng:2002fv}
  B.~Feng, S.~Franco, A.~Hanany and Y.~H.~He,
  %``Unhiggsing the del Pezzo,''
  JHEP {\bf 0308} (2003) 058
  [arXiv:hep-th/0209228].
  %%CITATION = JHEPA,0308,058;%%

\bibitem{Feng:2002zw}
  B.~Feng, S.~Franco, A.~Hanany and Y.~H.~He,
%  ``Symmetries of toric duality,''
  JHEP {\bf 0212}, 076 (2002)
  [arXiv:hep-th/0205144].
%27


\bibitem{master}
 D.~Forcella, A.~Hanany, Y.~H.~He and A.~Zaffaroni,
%  ``The Master Space of N=1 Gauge Theories,''
  JHEP {\bf 0808}, 012 (2008)
  [arXiv:0801.1585 [hep-th]]; 
%  `Mastering the Master Space,''
  Lett.\ Math.\ Phys.\  {\bf 85}, 163 (2008)
  [arXiv:0801.3477 [hep-th]].
    D.~Forcella,
%  ``Master Space and Hilbert Series for N=1 Field Theories,''
  arXiv:0902.2109 [hep-th].
  %28
 \bibitem{Imamura:2008qs}
  Y.~Imamura and K.~Kimura,
%  ``Quiver Chern-Simons theories and crystals,''
  arXiv:0808.4155 [hep-th].
%29
\bibitem{fano3}
S.~Mori, S.~Mukai,
%``Classification of Fano 3-folds with $B_2 \ge 2$, I'', in Algebraic
%and topological theories Nagata, M. (ed.) et al., 
Papers from the symposium
dedicated to the memory of Dr. Takehiko Miyata held in Kinosaki, October
30-November 9, 1984. Tokyo: Kinokuniya Company Ltd. 496-545 (1986)\\
J.-P.~Murre, 
%``Classification of Fano threefolds according to Fano and Iskovskih,''
%in Algebraic threefolds, 
Proc. 2nd 1981 Sess. C.I.M.E., Varenna/Italy 1981,
Lect. Notes Math. 947, 35-92 (1982).\\
S.~D.~Cutkosky, 
%``On Fano 3-folds'', 
Manuscr. Math. 64, No.2, 189-204 (1989).
%30
  %\cite{Fabbri:1999hw}
\bibitem{Fabbri:1999hw}
  D.~Fabbri, P.~Fre', L.~Gualtieri, C.~Reina, A.~Tomasiello, A.~Zaffaroni and A.~Zampa,
%  ``3D superconformal theories from Sasakian seven-manifolds: New  nontrivial
  %evidences for AdS(4)/CFT(3),''
  Nucl.\ Phys.\  B {\bf 577}, 547 (2000)
  [arXiv:hep-th/9907219].
  %%CITATION = NUPHA,B577,547;%%  
%last
\bibitem{future}
J.~Davey, et al., ``M2-Branes and Fano 3-folds''. To appear  


\begin{comment}


\bibitem{Feng:2001xr}
  B.~Feng, A.~Hanany and Y.~H.~He,
  ``Phase structure of D-brane gauge theories and toric duality,''
  JHEP {\bf 0108}, 040 (2001)
  [arXiv:hep-th/0104259].



\bibitem{Feng:2002fv}
  B.~Feng, S.~Franco, A.~Hanany and Y.~H.~He,
  ``Unhiggsing the del Pezzo,''
  JHEP {\bf 0308}, 058 (2003)
  [arXiv:hep-th/0209228].
  
  \bibitem{Feng:2001bn}
  B.~Feng, A.~Hanany, Y.~H.~He and A.~M.~Uranga,
  ``Toric duality as Seiberg duality and brane diamonds,''
  JHEP {\bf 0112}, 035 (2001)
  [arXiv:hep-th/0109063].\\
   C.~E.~Beasley and M.~R.~Plesser,
  ``Toric duality is Seiberg duality,''
  JHEP {\bf 0112}, 001 (2001)
  [arXiv:hep-th/0109053].
  
%\cite{Franco:2003ea}
\bibitem{Franco:2003ea}
  S.~Franco, A.~Hanany and Y.~H.~He,
  ``A trio of dualities: Walls, trees and cascades,''
  Fortsch.\ Phys.\  {\bf 52}, 540 (2004)
  [arXiv:hep-th/0312222].
  %%CITATION = FPYKA,52,540;%%
    
%\cite{Franco:2003ja}
\bibitem{Franco:2003ja}
  S.~Franco, A.~Hanany, Y.~H.~He and P.~Kazakopoulos,
  ``Duality walls, duality trees and fractional branes,''
  arXiv:hep-th/0306092.
  %%CITATION = HEP-TH/0306092;%%

%\cite{Franco:2002mu}
\bibitem{Franco:2002mu}
  S.~Franco and A.~Hanany,
  ``Toric duality, Seiberg duality and Picard-Lefschetz transformations,''
  Fortsch.\ Phys.\  {\bf 51}, 738 (2003)
  [arXiv:hep-th/0212299].
  %%CITATION = FPYKA,51,738;%%

%\cite{Feng:2002kk}
\bibitem{Feng:2002kk}
  B.~Feng, A.~Hanany, Y.~H.~He and A.~Iqbal,
  ``Quiver theories, soliton spectra and Picard-Lefschetz transformations,''
  JHEP {\bf 0302}, 056 (2003)
  [arXiv:hep-th/0206152].
  %%CITATION = JHEPA,0302,056;%%

%\cite{Forcella:2008ng}
\bibitem{Forcella:2008ng}
  D.~Forcella, A.~Hanany and A.~Zaffaroni,
  ``Master Space, Hilbert Series and Seiberg Duality,''
  arXiv:0810.4519 [hep-th].
  %%CITATION = ARXIV:0810.4519;%%
  
  %%%%%%%%%%%%%%%%%%%%%%%%%%%%%%%%%%%% toric dual 2+1
  
 %\cite{Franco:2008um}
\bibitem{Franco:2008um}
  S.~Franco, A.~Hanany, J.~Park and D.~Rodriguez-Gomez,
  ``Towards M2-brane Theories for Generic Toric Singularities,''
  JHEP {\bf 0812}, 110 (2008)
  [arXiv:0809.3237 [hep-th]].
  %%CITATION = JHEPA,0812,110;%%


  
%\cite{Franco:2009sp}
\bibitem{Franco:2009sp}
  S.~Franco, I.~R.~Klebanov and D.~Rodriguez-Gomez,
  ``M2-branes on Orbifolds of the Cone over $Q^{1,1,1}$,''
  arXiv:0903.3231 [hep-th].
  %%CITATION = ARXIV:0903.3231;%%
  

  

  
  
 %%%%%%%%%%%%%%%%%%%%%%%%%%%%%%%%%% Crystal
%\cite{Lee:2006hw}
\bibitem{Lee:2006hw}
  S.~Lee,
  ``Superconformal field theories from crystal lattices,''
  Phys.\ Rev.\  D {\bf 75}, 101901 (2007)
  [arXiv:hep-th/0610204].
  %%CITATION = PHRVA,D75,101901;%%

\bibitem{Lee:2007kv}
  S.~Lee, S.~Lee and J.~Park,
  ``Toric AdS(4)/CFT(3) duals and M-theory crystals,''
  JHEP {\bf 0705}, 004 (2007)
  [arXiv:hep-th/0702120].
  
\bibitem{Kim:2007ic}
  S.~Kim, S.~Lee, S.~Lee and J.~Park,
  ``Abelian Gauge Theory on M2-brane and Toric Duality,''
  Nucl.\ Phys.\  B {\bf 797}, 340 (2008)
  [arXiv:0705.3540 [hep-th]].
  


%%%%%%%%%%%%%%%%%%%%%%%%%%%%%%%%%%% partial resolutions

%\cite{Beasley:1999uz}
\bibitem{Beasley:1999uz}
  C.~Beasley, B.~R.~Greene, C.~I.~Lazaroiu and M.~R.~Plesser,
  ``D3-branes on partial resolutions of abelian quotient singularities of Calabi-Yau threefolds,''
  Nucl.\ Phys.\  B {\bf 566}, 599 (2000)
  [arXiv:hep-th/9907186].
  %%CITATION = NUPHA,B566,599;%%
  
  %\cite{Park:1999ep}
\bibitem{Park:1999ep}
  J.~Park, R.~Rabadan and A.~M.~Uranga,
  ``Orientifolding the conifold,''
  Nucl.\ Phys.\  B {\bf 570}, 38 (2000)
  [arXiv:hep-th/9907086].
  %%CITATION = NUPHA,B570,38;%%
  
 %%%%%%%%%%%%%%%%%%%%%%%%%%%%%%%%%%%%
 %\cite{Gubser:1998vd}
\bibitem{Gubser:1998vd}
  S.~S.~Gubser,
  ``Einstein manifolds and conformal field theories,''
  Phys.\ Rev.\  D {\bf 59}, 025006 (1999)
  [arXiv:hep-th/9807164].
  %%CITATION = PHRVA,D59,025006;%% 
  
\bibitem{Butti:2005vn}
  A.~Butti and A.~Zaffaroni,
  ``R-charges from toric diagrams and the equivalence of a-maximization and
  Z-minimization,''
  JHEP {\bf 0511}, 019 (2005)
  [arXiv:hep-th/0506232].
  %%CITATION = JHEPA,0511,019;%%

  
%%%%%%%%%%%%%%%%%%%%%%%%%%%%%%%%%%%%  

\bibitem{pleth}
  S.~Benvenuti, B.~Feng, A.~Hanany and Y.~H.~He,
  ``Counting BPS operators in gauge theories: Quivers, syzygies and plethystics,''
  JHEP {\bf 0711}, 050 (2007)
  [arXiv:hep-th/0608050].
  
%\bibitem{Hanany:2006uc}
  A.~Hanany and C.~Romelsberger,
  ``Counting BPS operators in the chiral ring of N = 2 supersymmetric gauge
  theories or N = 2 braine surgery,''
  Adv.\ Theor.\ Math.\ Phys.\  {\bf 11}, 1091 (2007)
  [arXiv:hep-th/0611346].
  %%CITATION = 00203,11,1091;%%
  
  B.~Feng, A.~Hanany and Y.~H.~He,
  ``Counting gauge invariants: The plethystic program,''
  JHEP {\bf 0703}, 090 (2007)
  [arXiv:hep-th/0701063].
  
  D.~Forcella, A.~Hanany and A.~Zaffaroni,
  ``Baryonic generating functions,''
  JHEP {\bf 0712}, 022 (2007)
  [arXiv:hep-th/0701236].
  
  J.~Gray, A.~Hanany, Y.~H.~He, V.~Jejjala and N.~Mekareeya,
  ``SQCD: A Geometric Apercu,''
  JHEP {\bf 0805}, 099 (2008)
  [arXiv:0803.4257 [hep-th]].
  
   A.~Hanany and N.~Mekareeya,
  ``Counting Gauge Invariant Operators in SQCD with Classical Gauge Groups,''
  JHEP {\bf 0810}, 012 (2008)
  [arXiv:0805.3728 [hep-th]].
  
  A.~Hanany, N.~Mekareeya and G.~Torri,
  ``The Hilbert Series of Adjoint SQCD,''
  arXiv:0812.2315 [hep-th].

\bibitem{Hanany:2008qc}
  A.~Hanany, N.~Mekareeya and A.~Zaffaroni,
  ``Partition Functions for Membrane Theories,''
  JHEP {\bf 0809}, 090 (2008)
  [arXiv:0806.4212 [hep-th]].
  
  %\cite{Kim:2009wb}
\bibitem{Kim:2009wb}
  S.~Kim,
  ``The complete superconformal index for N=6 Chern-Simons theory,''
  arXiv:0903.4172 [hep-th].
  %%CITATION = ARXIV:0903.4172;%%

  \bibitem{Butti:2007jv}
  A.~Butti, D.~Forcella, A.~Hanany, D.~Vegh and A.~Zaffaroni,
  ``Counting Chiral Operators in Quiver Gauge Theories,''
  JHEP {\bf 0711}, 092 (2007)
  [arXiv:0705.2771 [hep-th]].
  
%\cite{Benvenuti:2004dw}
\bibitem{Benvenuti:2004dw}
  S.~Benvenuti and A.~Hanany,
  ``New results on superconformal quivers,''
  JHEP {\bf 0604}, 032 (2006)
  [arXiv:hep-th/0411262].
  %%CITATION = JHEPA,0604,032;%%

\bibitem{Franco:2006gc}
  S.~Franco and D.~Vegh,
  ``Moduli spaces of gauge theories from dimer models: Proof of the
  correspondence,''
  JHEP {\bf 0611} (2006) 054
  [arXiv:hep-th/0601063].
  
\bibitem{Davide}
  A.~Amariti, D.~Forcella, L.~Girardello and A.~Mariotti,
  ``3D Seiberg-like Dualities and M2 Branes,''
  arXiv:0903.3222 [hep-th].

\bibitem{Seba}
S.~Franco, I.~Klebanov and D.~Rodriguez-Gomez,
``M2-branes on Orbifolds of the Cone over $Q^{1,1,1}$,''
arXiv:0903.3231 [hep-th].

\bibitem{Bergman:1999na}
  O.~Bergman, A.~Hanany, A.~Karch and B.~Kol,
  ``Branes and supersymmetry breaking in 3D gauge theories,''
  JHEP {\bf 9910}, 036 (1999)
  [arXiv:hep-th/9908075].
  %%CITATION = JHEPA,9910,036;%%
  
%\cite{Aganagic:2001ug}
\bibitem{Aganagic:2001ug}
  M.~Aganagic and C.~Vafa,
  ``G(2) manifolds, mirror symmetry and geometric engineering,''
  arXiv:hep-th/0110171.
  %%CITATION = HEP-TH/0110171;%%
  
%\cite{Imamura:2009ur}
\bibitem{Imamura:2009ur}
  Y.~Imamura,
  ``Monopole operators in N=4 Chern-Simons theories and wrapped M2-branes,''
  arXiv:0902.4173 [hep-th].
  %%CITATION = ARXIV:0902.4173;%%
  
  %\cite{Imamura:2008nn}
\bibitem{Imamura:2008nn}
  Y.~Imamura and K.~Kimura,
  ``On the moduli space of elliptic Maxwell-Chern-Simons theories,''
  Prog.\ Theor.\ Phys.\  {\bf 120}, 509 (2008)
  [arXiv:0806.3727 [hep-th]].
  %%CITATION = PTPKA,120,509;%%
  
  %\cite{Imamura:2008ji}
\bibitem{Imamura:2008ji}
  Y.~Imamura and S.~Yokoyama,
  ``N=4 Chern-Simons theories and wrapped M5-branes in their gravity duals,''
  arXiv:0812.1331 [hep-th].
  %%CITATION = ARXIV:0812.1331;%%
  

%%%%%%%%%%%%%%%%%%%%%%%%%%%%%%%%%
 %\cite{Aganagic:2009zk}
\bibitem{Aganagic:2009zk}
  M.~Aganagic,
  ``A Stringy Origin of M2 Brane Chern-Simons Theories,''
  arXiv:0905.3415 [hep-th].
  %%CITATION = ARXIV:0905.3415;%% 

\bibitem{ucsbtalk}
A.~Hanany,
``Finding M Theory Duals to Type IIA Backgrounds with RR Fluxes''
(talk at \emph{Fundamental Aspects of Superstring Theory}),
May 29, 2009, Kavli Institute for Theoretical Physics (KITP): 
{\sf http://online.itp.ucsb.edu/online/strings09/hanany2/}
%%%%%%%%%%%%%%%%%%%%%%%%%%%%%%%%%  
  
%\cite{Franco:2002ae}
\bibitem{Franco:2002ae}
  S.~Franco and A.~Hanany,
  ``Geometric dualities in 4d field theories and their 5d interpretation,''
  JHEP {\bf 0304}, 043 (2003)
  [arXiv:hep-th/0207006].
  %%CITATION = JHEPA,0304,043;%%  
  
 %\cite{Hanany:2001py}
\bibitem{Hanany:2001py}
  A.~Hanany and A.~Iqbal,
  ``Quiver theories from D6-branes via mirror symmetry,''
  JHEP {\bf 0204}, 009 (2002)
  [arXiv:hep-th/0108137].
  %%CITATION = JHEPA,0204,009;%%
  
 %\cite{Uranga:1998vf}
\bibitem{Uranga:1998vf}
  A.~M.~Uranga,
  ``Brane Configurations for Branes at Conifolds,''
  JHEP {\bf 9901}, 022 (1999)
  [arXiv:hep-th/9811004].
  %%CITATION = JHEPA,9901,022;%% 
  
 %\cite{Erlich:1999rb}
\bibitem{Erlich:1999rb}
  J.~Erlich, A.~Hanany and A.~Naqvi,
  ``Marginal deformations from branes,''
  JHEP {\bf 9903}, 008 (1999)
  [arXiv:hep-th/9902118].
  %%CITATION = JHEPA,9903,008;%% 
  
%%%%%%%%%%%%%%%%%% fano


\bibitem{Aharony:2008ug}
  O.~Aharony, O.~Bergman, D.~L.~Jafferis and J.~Maldacena,
 % ``N=6 superconformal Chern-Simons-matter theories, M2-branes and their
  %gravity duals,''
  arXiv:0806.1218 [hep-th].
%%%%%%%%%%%%%%%%%%%%%%%%%%%%%%%%%%%%%  
   %\cite{Petrini:2009ur}
   \bibitem{Petrini:2009ur}
  M.~Petrini and A.~Zaffaroni,
 % ``N=2 solutions of massive type IIA and their Chern-Simons duals,''
  arXiv:0904.4915 [hep-th].
  %%CITATION = ARXIV:0904.4915;%%
  


\end{comment}
}

\end{thebibliography}
\end{document}